\documentclass{ws-ijmpc}

\begin{document}

\markboth{Dilip P. Ahalpara and Jitendra C. Parikh}
{MODELING TIME SERIES DATA OF REAL SYSTEMS (Paper's Title)}

\catchline{}{}{}{}{}

\title{MODELING TIME SERIES DATA OF REAL SYSTEMS}

\author{DILIP P. AHALPARA}

\address{Institute for Plasma Research, Near Indira Bridge, Bhat\\
Gandhinagar-382428, India\\
dilip@ipr.res.in}

\author{JITENDRA C. PARIKH}

\address{Physical Research Laboratory, Navrangpura\\
Ahmedabad-380009, India\\
parikh@prl.res.in}

\maketitle

\begin{abstract}
Dynamics of complex systems is studied by first considering a chaotic time series generated by Lorenz equations and adding noise to it. The trend (smooth behavior) is separated from fluctuations at different scales using wavelet analysis and a prediction method proposed by Lorenz is applied to make out of sample predictions at different regions of the time series. The prediction capability of this method is studied by considering several improvements over this method. We then apply this approach to a real financial time series. The smooth time series is modeled using techniques of non linear dynamics. Our results for predictions suggest that the modified Lorenz method gives better predictions compared to those from the original Lorenz method. Fluctuations are analyzed using probabilistic considerations.

\end{abstract}

\ccode{PACS Nos.: 05.45.Tp, 02.30.NW}

\section{Introduction}

Dynamics of a real complex system (in physics, geology, physiology, finance) is often inferred/ constructed by studying and analyzing time series data of relevant system variables [1,2,3]. In such systems, dynamical behavior is many times characterized by features having vastly different time scales. Thus, we often observe in the time series data, long term trends (mean behavior) superimposed by rapid variations (fluctuations) at much shorter time scales. In addition, the data may also contain high frequency noise. Consequently, it seems appropriate to separate the trend from the fluctuating component in a time series. For a time series $X_i$, i=1, 2, ... n, this separation may be explicitly written as

\begin{equation}
X_i = X_i(m) + X_i(f), \ \ \   i=1, 2, ... n
\label{eqn:MeanFluct}
\end{equation}

where $X_i(m)$ and $X_i(f)$ are the mean and the fluctuating components respectively. As in statistical physics, we therefore model trend using a deterministic framework and fluctuations using a stochastic framework.

To illustrate our approach, we first consider the time series (X-component) of the chaotic Lorenz model with and without addition of noise. In addition, we also apply our method to a real financial (NASDAQ composite Index) time series.

Clearly, the first step in working out this program is to separate average behavior from fluctuations in the time series. We use discrete wavelets [4] to carry out the separation and obtain a time series of the trends. This is described in Sec. 2.

In order to model the trend (time) series, standard techniques of nonlinear dynamics are employed. More precisely, using methods of delays an embedding is constructed that describes the dynamics of the system on an attractor. A brief discussion of the method and our results for the two time series are given in Sec. 3.

We have generated predictions using a method proposed by Lorenz and modified by us. Sec. 4 describes in brief the method and gives a comparison of predictions using the two. Sec. 5 contains analysis of fluctuations in the financial time series. Finally, in Sec. 6, we present a brief summary and discuss our results.

\section{Smoothening the Time Series Data Set}

A number of methods exist in the literature and have been used for separating fluctuations from trend in a time series. In this context, it is important to point out that most time series of real systems with complex dynamics are non-stationary in nature. Consequently in such cases, we need to employ a suitable method to separate the fluctuations from the trend. The method of de-trended fluctuation analysis (DFA) [5] and its extension multi-fractal DFA (MFDFA) [6] use local polynomial (linear, quadratic) fits to describe the trend. We use discrete wavelet transforms (DWT) [4] for this purpose. Such an approach has recently been proposed by Manimaran et al [7].

The basic reason for our choice of DWT is related to their nice mathematical properties. In the present context, it is sufficient to note that (i) DWT provides a complete orthonormal basis to decompose a non-stationary  signal and (ii) wavelet functions have a finite number of moments that are zero. In our work we have used length-4 Daubechies-4 (Db-4) wavelet transform. It is one of the simplest and smallest (even) length wavelet transform that is smooth. We next describe in brief our wavelet based procedure.

Given a time series composed of n points, namely $X_i$, i=1, 2, ... n, we first carry out a forward Db-4 wavelet transformation [4] that gives n wavelet coefficients. Of these, half (n/2) are low pass coefficients that describe the average behavior locally and the other half (n/2) are high pass coefficients corresponding to local fluctuations. In order to obtain a smooth time series, the high pass coefficients are set to zero and then an inverse Db-4 transformation is carried out. This results in smoothening of the data set (see Manimaran et al [7]). While using the Db-4 transform it has been observed that due to fixed boundary of the data set, rapid fluctuations are observed towards the beginning and the end of the smoothened data set. In order to remove this spurious effect, we do a padding of the data set by adding constant valued n/2 data points at the beginning and n/2 data points at the end of the time series. The constant value matches with the value of the first and the last data point respectively. The forward Db-4 transformation is then applied on the data set having 2n data points. Having smoothened the data set by an inverse Db-4 transformation, the padded data sets (containing the spurious effect) are removed thereby getting an improved smoothened time series of the n data set points. We thus obtain a {\em level one} time series of trends in which fluctuations at the smallest scale have been filtered out and the trend extracted after applying Db-4 transform. One can repeat the entire process on {\em level one} series to filter out fluctuations at the next higher time scale to get a {\em level two} smoothened series and so on. It is worth emphasizing that the low-pass wavelet coefficients give a representation in the transformed space of the smooth determistic part of the time series.

We have used Lorenz X-series (having 8192 data points) with and without noise and NASDAQ composite index time series (having 2048 data points) and have carried out smoothening at various levels for the NASDAQ data. The daily closing value of the index used by us covers the period 20-Feb-1996 to 7-Apr-2004.

We consider the time series of variable X obtained by solving Lorenz equations given below:

\begin{equation}
\dot{X} = \sigma(Y - X), \ \   \dot{Y} = \rho X - Y - XZ, \ \     \dot{Z} = - \beta Z + XY
\label{eqn:Lorenz}
\end{equation}

The parameters that we have used in the solution of these equations are $\sigma = 16.0$, $\rho = 45.92$ and  $\beta = 4.0$. The initial 5000 points generated in the series are omitted to get rid of the transient effects. Subsequently the series of X variable has been calculated by integrating the equations using Runge-Kutta method (RK4). A small time step of 0.001 has been used for obtaining a better accuracy. However the actual time series used in the present analysis was obtained by using the sampling interval of 0.05 by selecting every $50^{th}$ value from the calculated series. Thus we obtained a Lorenz's time series of 8192 values.

Figs. 1 (a) to (d) show the Lorenz X-series (showing 4 regions to be discussed later). Here we also consider the series obtained by adding noise [8] having a Gaussian distribution to the Lorenz X-series. The Gaussian noise being considered has a zero mean and a standard deviation that is equal to some percentage of the standard deviation of the signal into which it is to be superimposed. More precisely, we generated the new series having Gaussian noise as follows:

\begin{equation}
Y_i = X_i + \alpha Z_i \ \ \    (i = 1, 2, ... \ N)
\label{eqn:GaussianNoise}
\end{equation}

where set $\{X_i\}$ is the original Lorenz series and the set $\{Z_i\}$ represents N independent Gaussian variates with mean $<Z>$ = 0 and variance $\sigma^{2}=1.0$. Here parameter $\alpha \equiv k (\sigma_x / \sigma_z) = k \sigma_x$, where $\sigma_x^{2}$ is the variance of the time series $\{X_{i}\}$. Further, k in per cent denotes the amount of noise we add to the original pure (i.e. noise free) time series $\{X_{i}\}$. We have considered different percentages of noise strength k and analyzed these time series and present here results obtained for k=1\%, 5\% and 10\% noise level. The Lorenz X-series added with Gaussian noise is first smoothened using Db-4 level 1 transformation. In Fig. 1 we show pure Lorenz X-series (Figs. (a) to (d) where the arrows mark 4 regions that we consider for predictions which we discuss later (Sec. 4)). In order to highlight the smoothening effects, the figure also shows a comparison of the actual and the smoothened data sets (the X-series added with 5\% Gaussian noise and the smoothened level 1 X-series) at the beginning (Fig. 1 (e)) and at the end (Fig. 1 (f)), 50 points each of the time series.

\begin{figure}
\label{fig:Lorenz}
\centering
  \begin{tabular}{ccccc}
  \epsfig{file=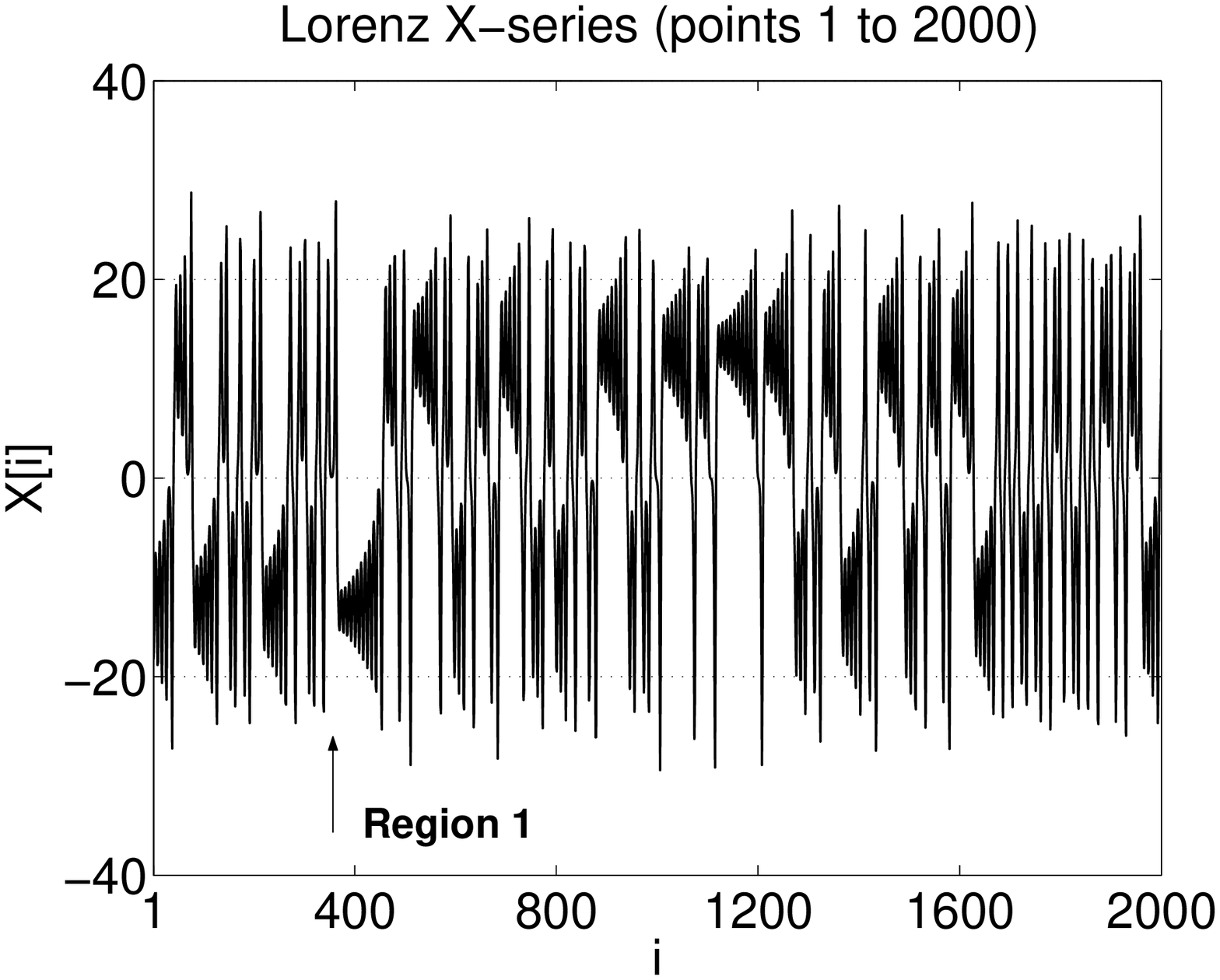,width=0.22\linewidth,clip=} &
  \epsfig{file=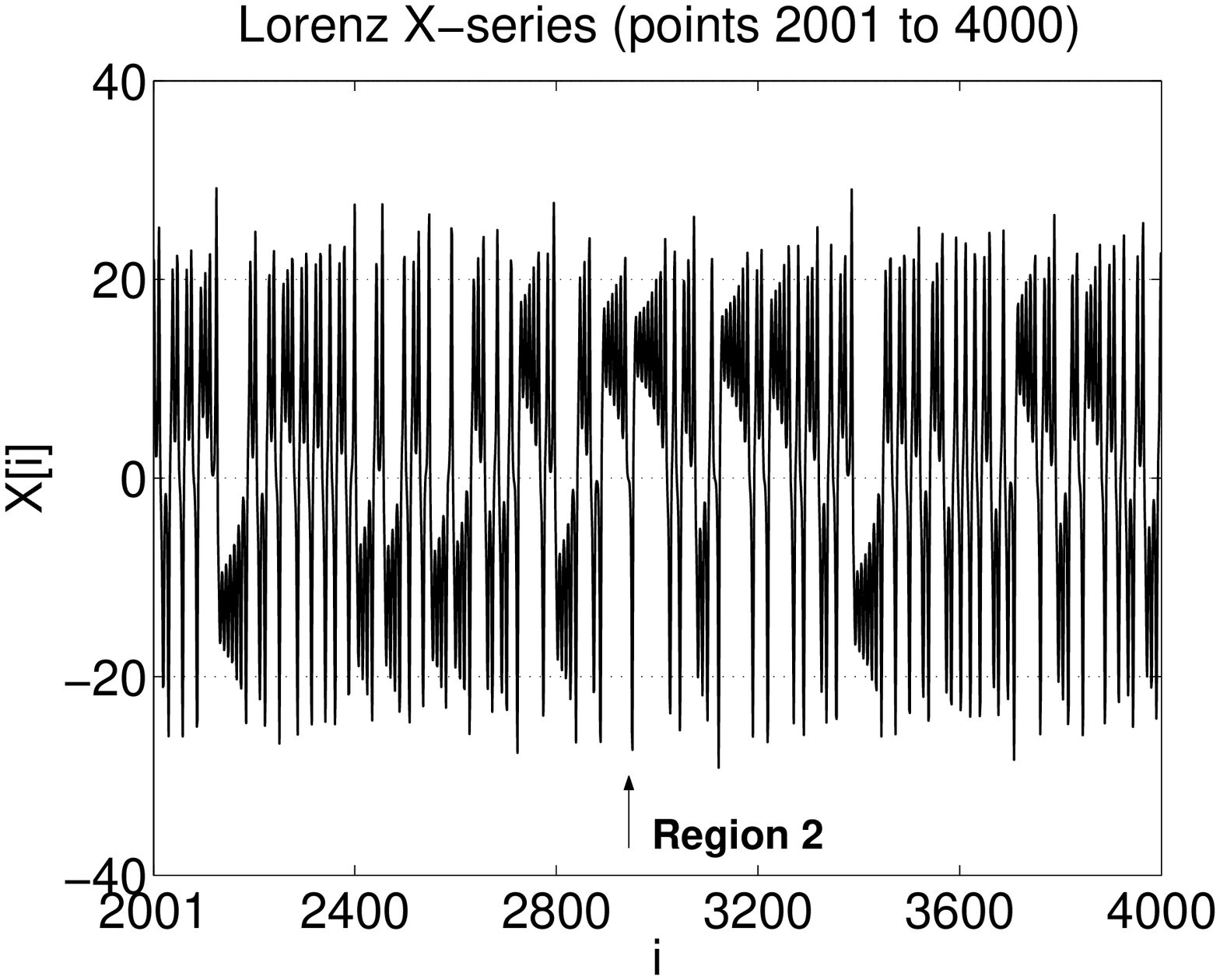,width=0.22\linewidth,clip=} &
  \epsfig{file=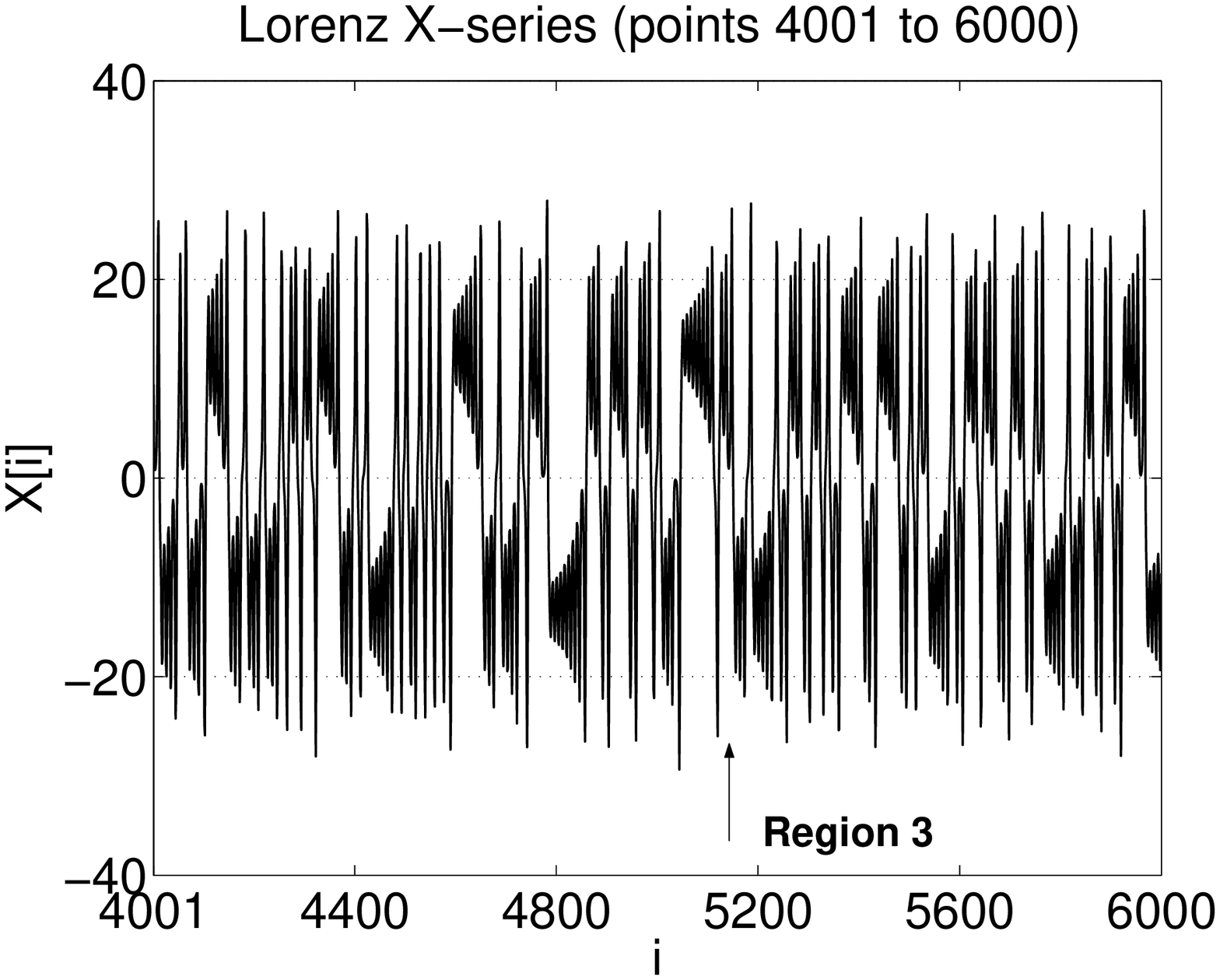,width=0.22\linewidth,clip=} &
  \epsfig{file=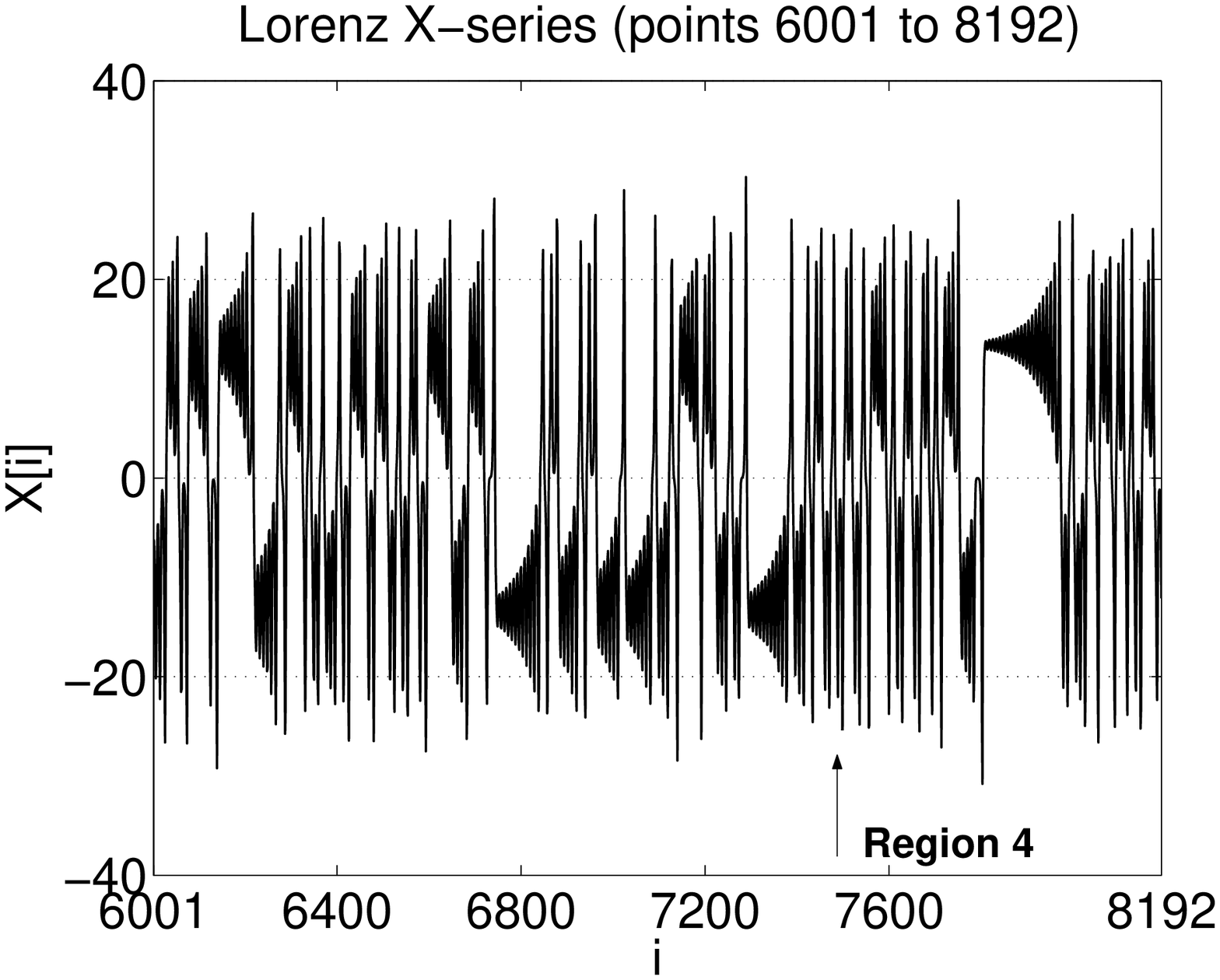,width=0.22\linewidth,clip=} \\
  \end{tabular}
  \vspace{-0.18in}
    \begin{verbatim}        (a)              (b)               (c)              (d)\end{verbatim}

  \begin{tabular}{cc}
  \epsfig{file=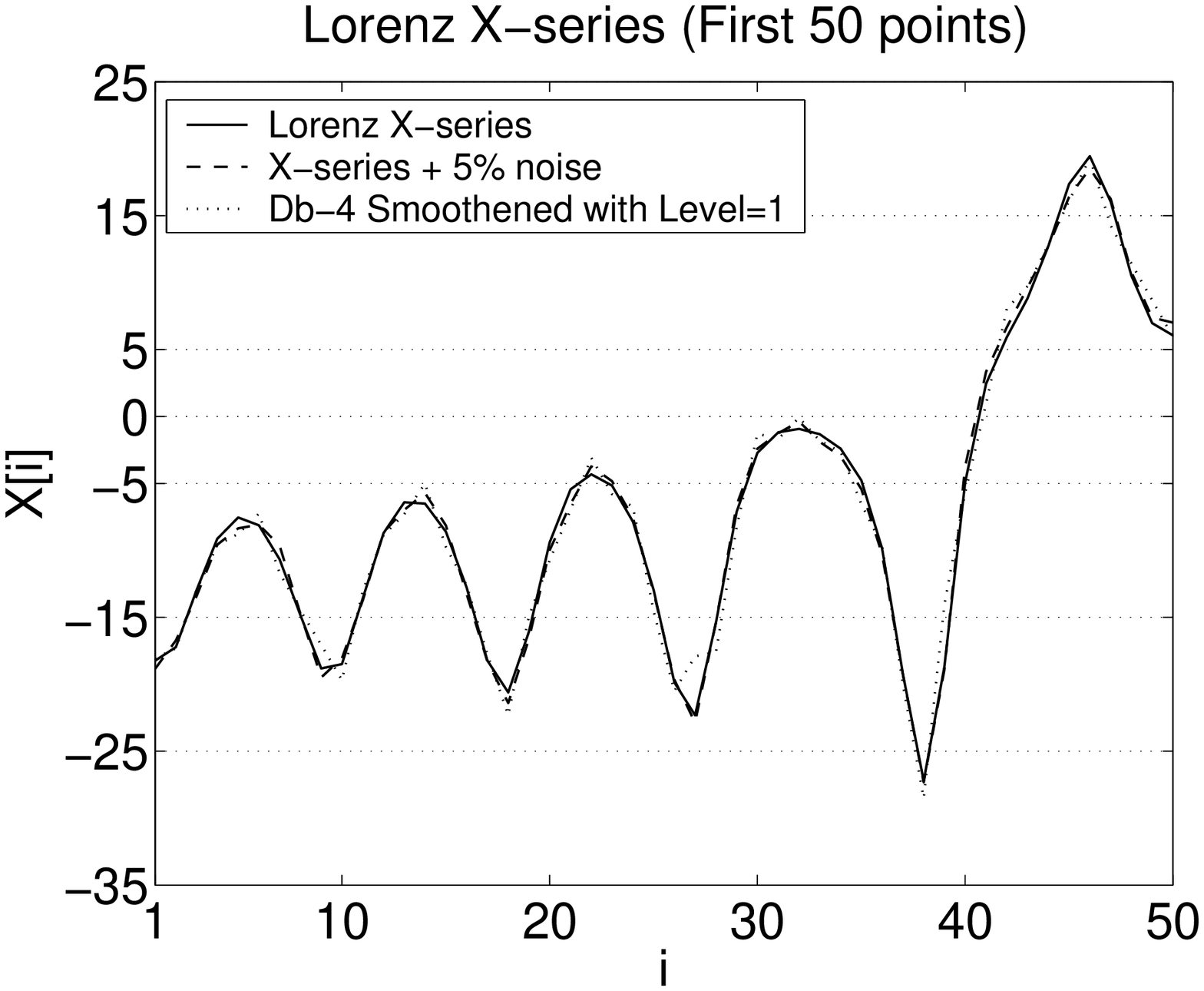,width=0.40\linewidth,clip=} & 
  \epsfig{file=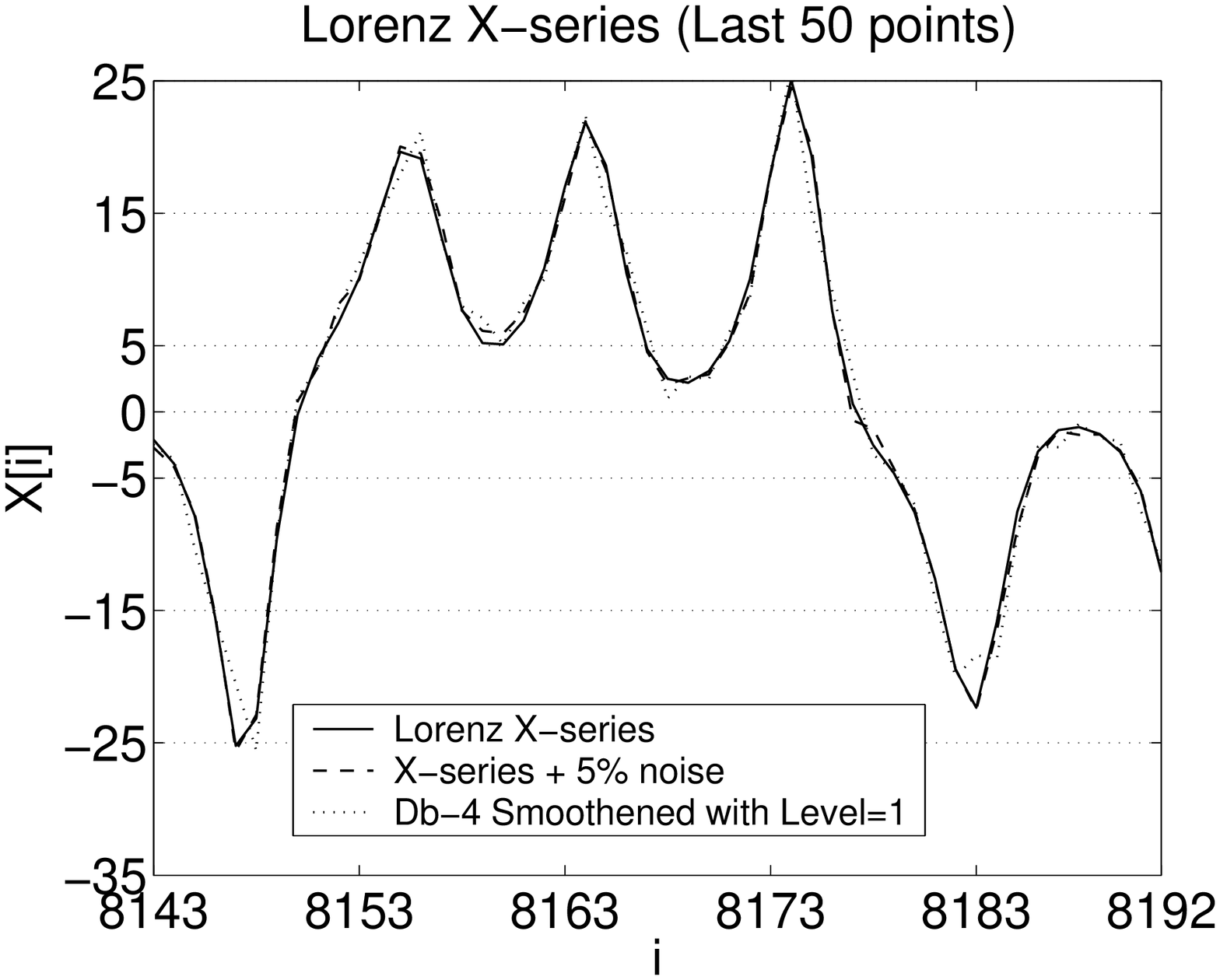,width=0.40\linewidth,clip=}
  \end{tabular}
  \vspace{-0.15in}
    \begin{verbatim}                   (e)                           (f)\end{verbatim}

\vspace{-0.15in}
\caption{The Lorenz X-series having 8192 points (Figs. (a) to (d)). The first and last 50 points of X-series (Figs. (e) and (f) respectively) are compared with addition of 5\% Gaussian noise and then smoothening it by Db-4 transform with Level=1.}
\end{figure}

Fig. 2 (a) shows the original NASDAQ composite index time series and the smoothened time series at levels 1 and 4. In order to highlight the smoothening effects, the figure 2 also contains a comparison of the actual and the smoothened data sets at the beginning (Fig. 2 (b)) and at the end (Fig. 2 (c)), 100 points each of the time series. Note that the level 4 time series has filtered out fluctuations over a larger 32 time steps window size retaining only the broad trends in the original series.

\begin{figure}
\centering
  \begin{tabular}{c}
  \epsfig{file=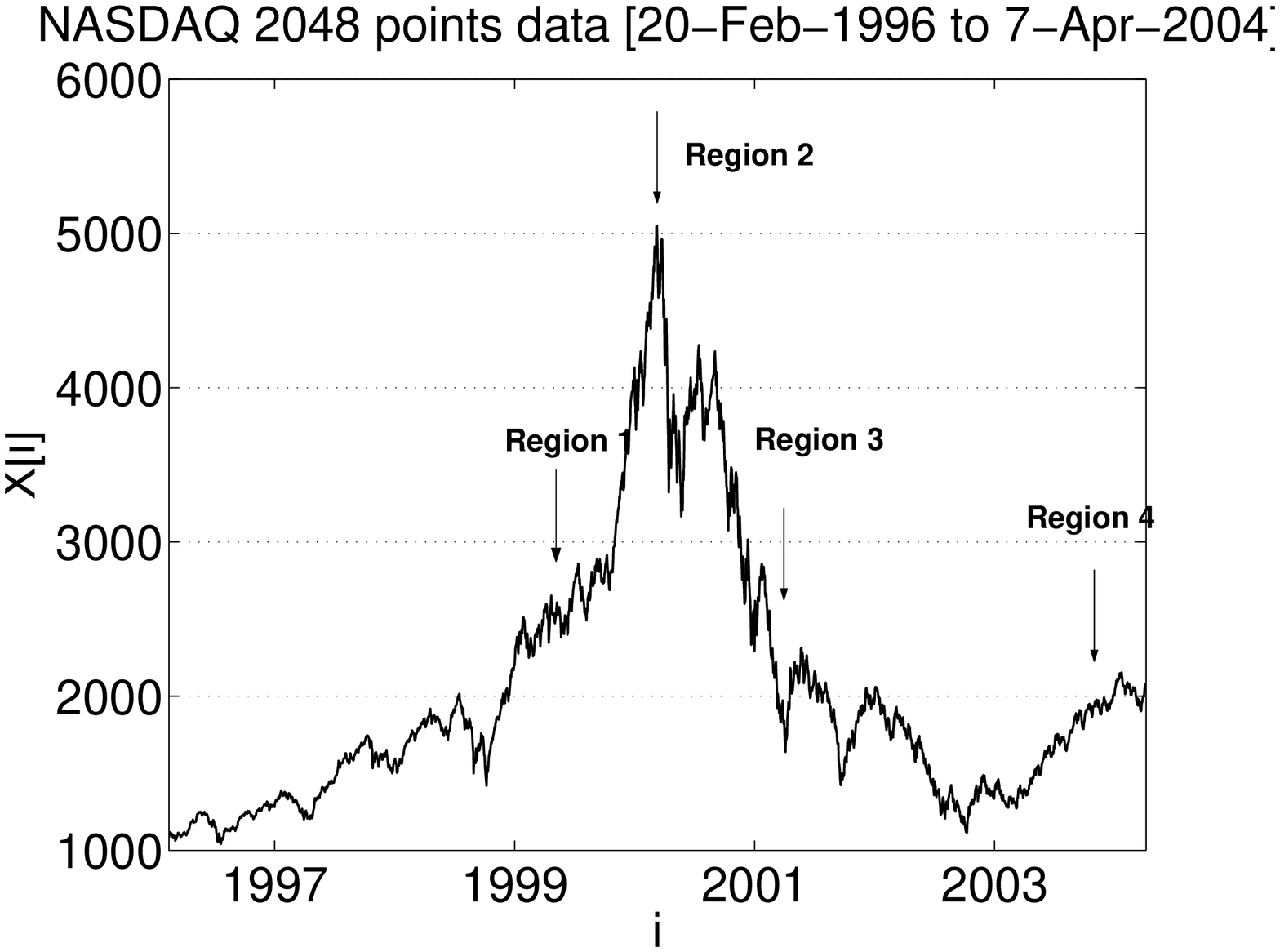,width=0.40\linewidth,clip=} \\ 
  \end{tabular}
  \vspace{-0.15in}
    \begin{verbatim}                                 (a)\end{verbatim}

  \begin{tabular}{cc}
  \epsfig{file=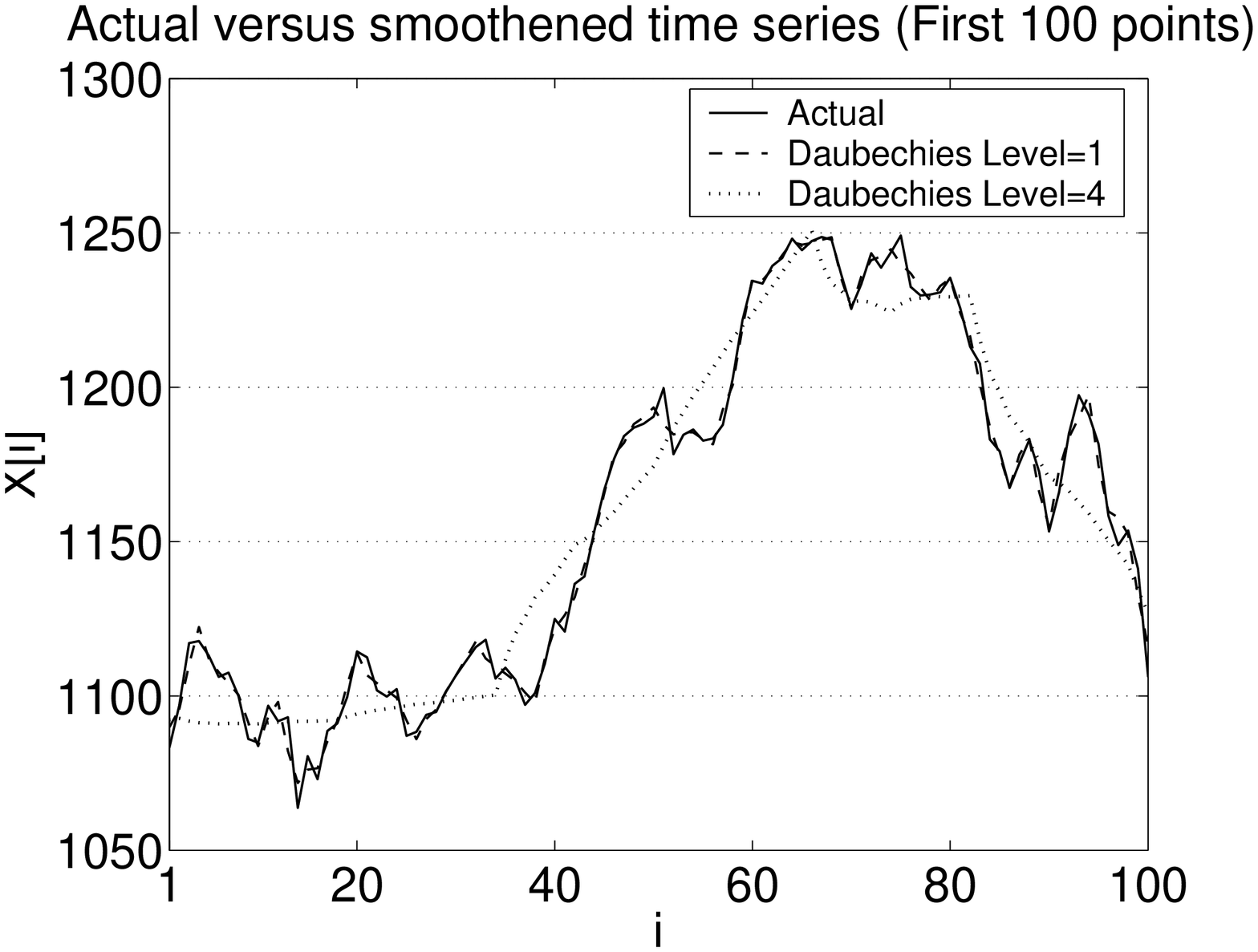,width=0.40\linewidth,clip=} & 
  \epsfig{file=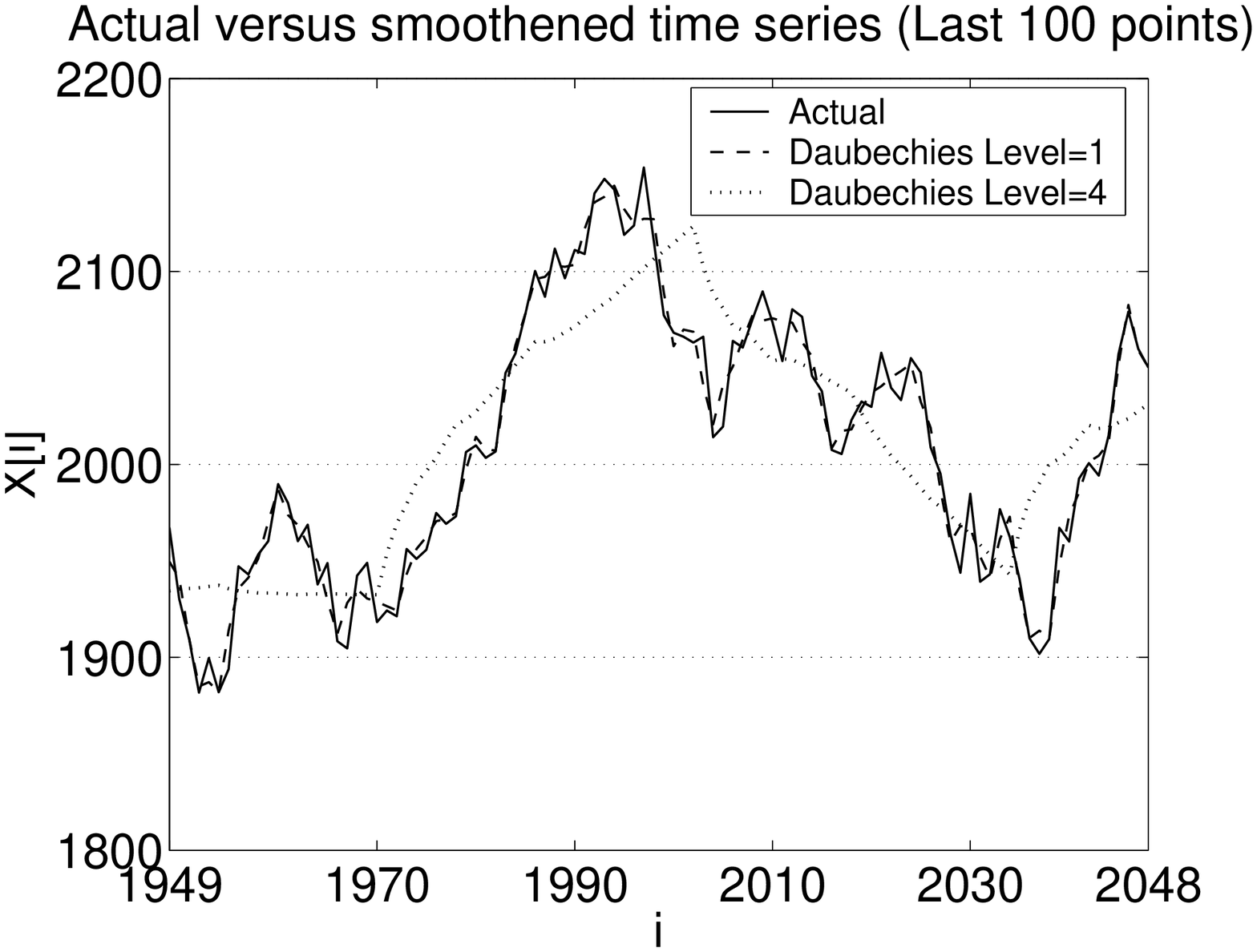,width=0.40\linewidth,clip=} \\
  \end{tabular}
  \vspace{-0.15in}
    \begin{verbatim}                  (b)                           (c)\end{verbatim}

\vspace{-0.15in}
\caption{(a) The actual time series of 2048 points (NASDAQ composite index) compared with the smoothened Db-4 time series at levels 1 and 4. In (b) and (c) the time series shows the first and the last 100 points respectively.}
\end{figure}

We show in Fig. 3 the low pass wavelet coefficients at level 1 and level 4 for the NASDAQ composite index.

\begin{figure}
\centering

  \begin{tabular}{cc}
  \epsfig{file=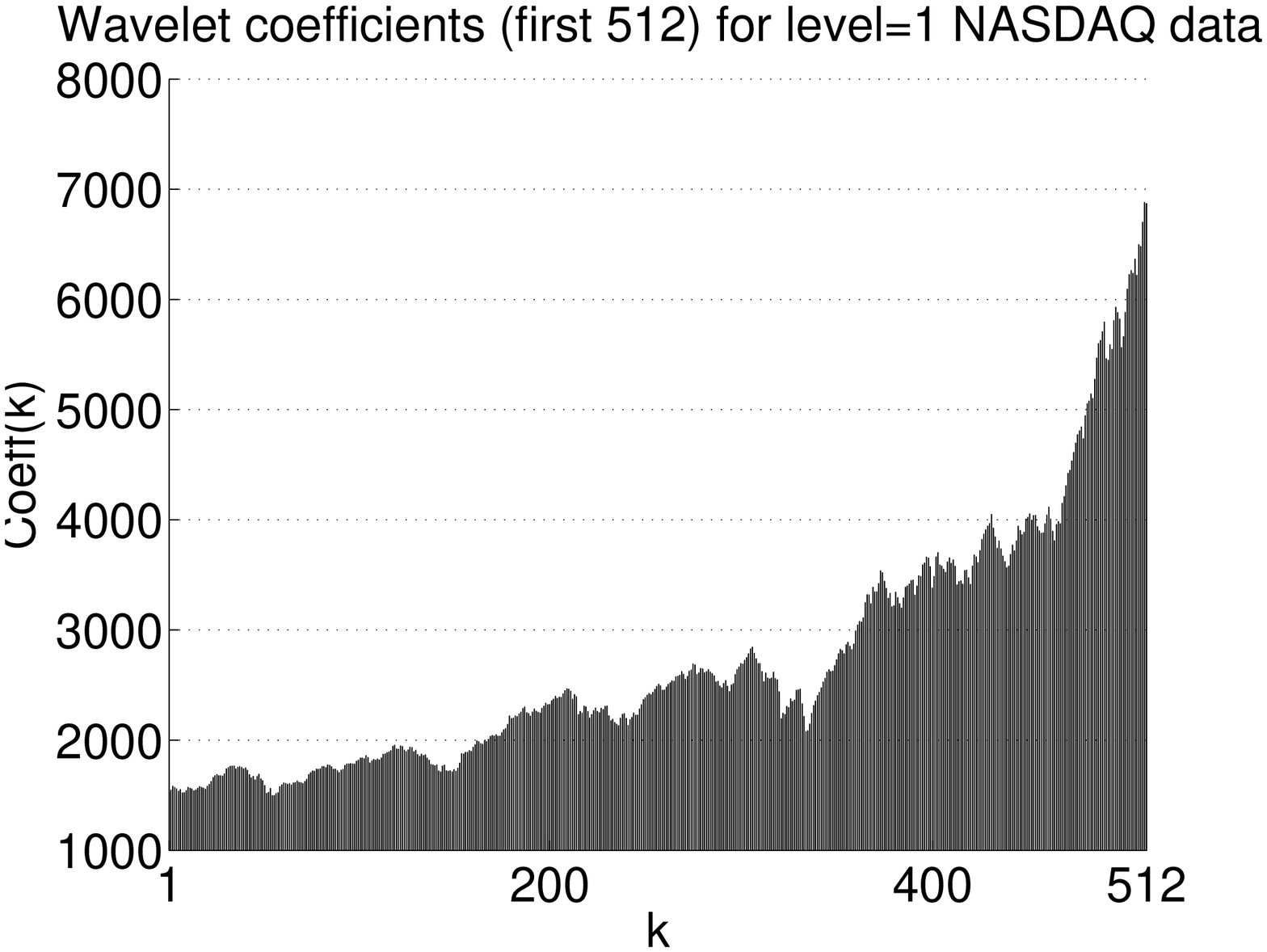,width=0.40\linewidth,clip=} & 
  \epsfig{file=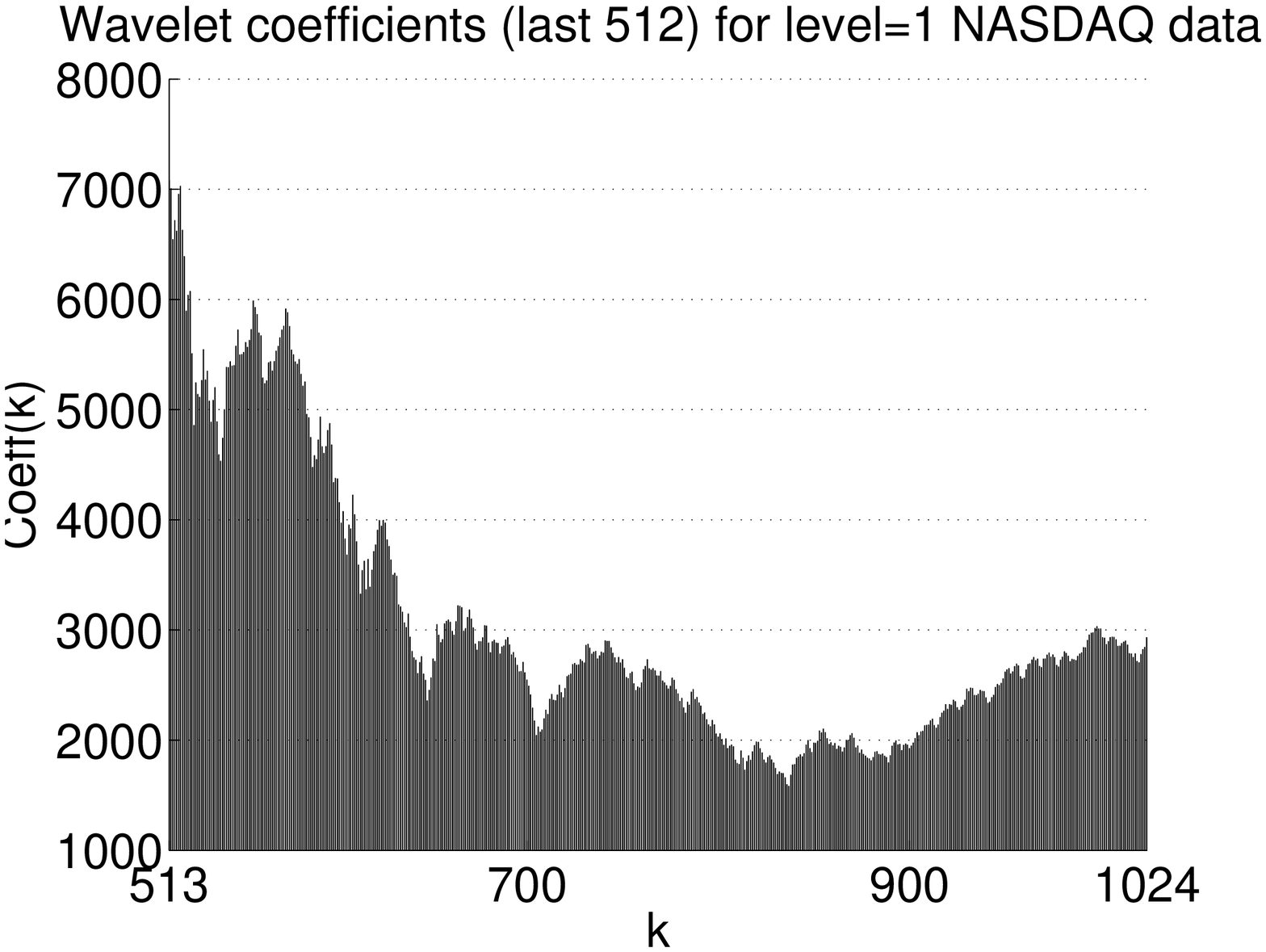,width=0.40\linewidth,clip=} \\
  \end{tabular}
  \vspace{-0.15in}
    \begin{verbatim}                  (a)                           (b)\end{verbatim}

  \begin{tabular}{c}
  \epsfig{file=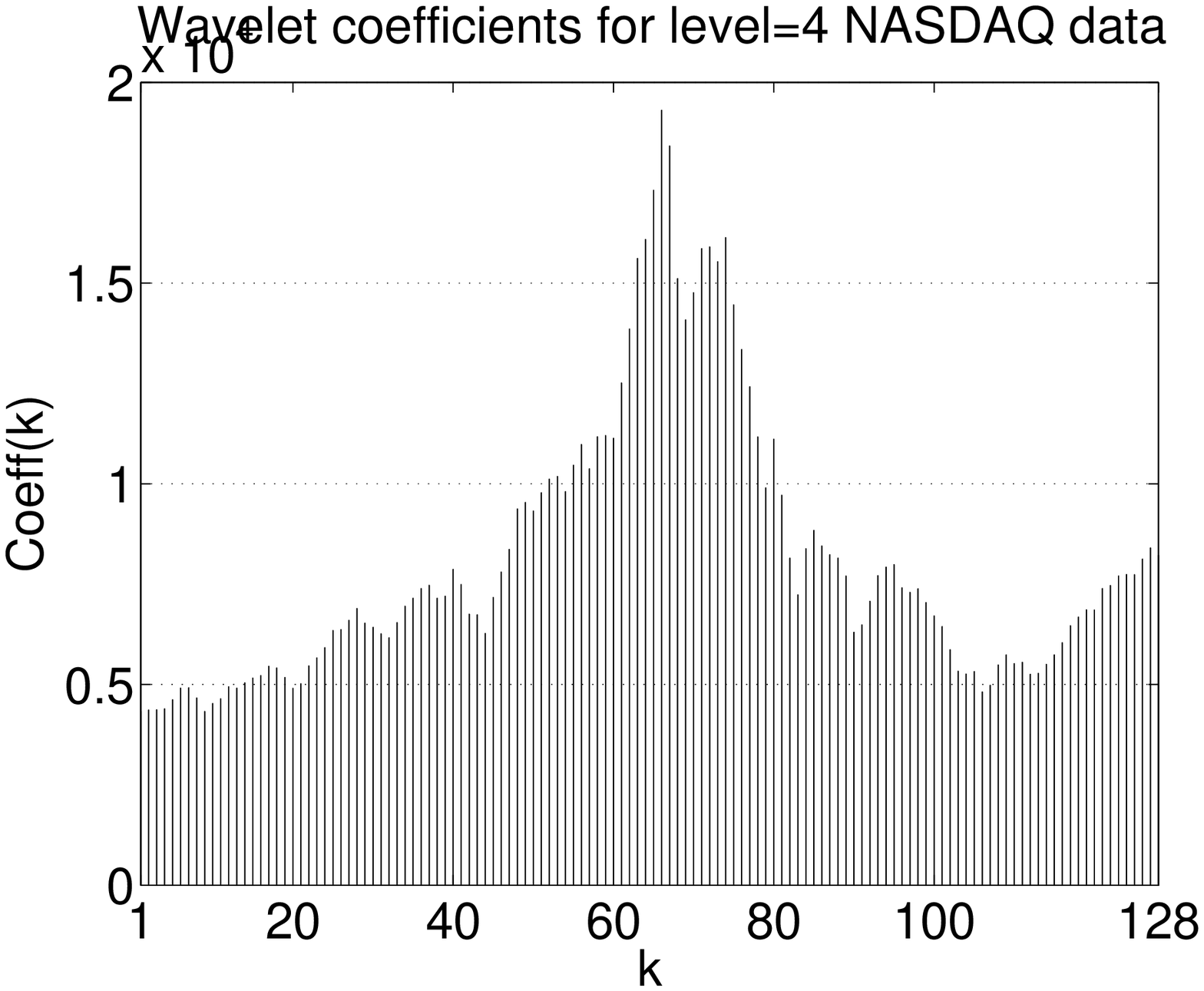,width=0.40\linewidth,clip=} \\ 
  \end{tabular}
  \vspace{-0.15in}
    \begin{verbatim}                                  (c)\end{verbatim}

\vspace{-0.15in}
\caption{The low pass wavelet coefficients for level 1 (1024 coefficients are shown in (a) and (b) corresponding to first 512 coefficients and last 512 coefficients respectively) and level 4 (128 coefficients are shown in (c)) for the 2048 points data set of NASDAQ composite index.}
\end{figure}

As expected the low pass coefficients at level 1 exhibit more structure than the ones at level 4. Further, they show a pattern which is similar to the original time series.

\section{State Space Reconstruction}

All the smoothened time series are assumed to describe deterministic dynamics and we attempt to model them in the framework of state space reconstruction with time delayed vectors.

We first use the standard time delay embedding approach as a means of reconstructing the vector space that is equivalent to the original state space of the system using the smoothened time series. In order to carry out the embedding we need to determine 2 parameters, namely time delay $\tau$ and embedding dimension d.

Average mutual information analysis is used to obtain the time delay $\tau$ and the number of false nearest neighbors analysis is used to obtain the embedding dimension d. These two methods are described by Abarbanel et al [9].

\subsection{Lorenz X-series}

We first consider pure Lorenz X-series. The correlations in pure Lorenz X-series fall off rapidly (Fig. 4 (a)). We use the prescription $I(\tau)/I(0) \approx 0.2$ suggested by Abarbanel et al [9] for choosing the time delay $\tau=1$.  Here $I(T)$ represents average mutual information as a function of time lag T. Since this is a prescription, we also allow for a variation in value of $\tau$, as suggested by Kulkarni et al [8]. In the present case, we consider $\tau=1$ as well as $\tau=2$. The dimension is fixed by choosing the smallest dimension for which number of false neighbors become zero. Further we require that the number of false neighbors consistently remains zero there after for higher dimensions. We have used this criterion for all the time series considered in our analysis. Accordingly it is seen from Fig. 4 (b) and Table 1 that for pure Lorenz X-series with $\tau=1$ we get $d=3$. Table 1 gives the values obtained for false neighbors for this time series.

\begin{figure}
\centering

  \begin{tabular}{cc}
  \epsfig{file=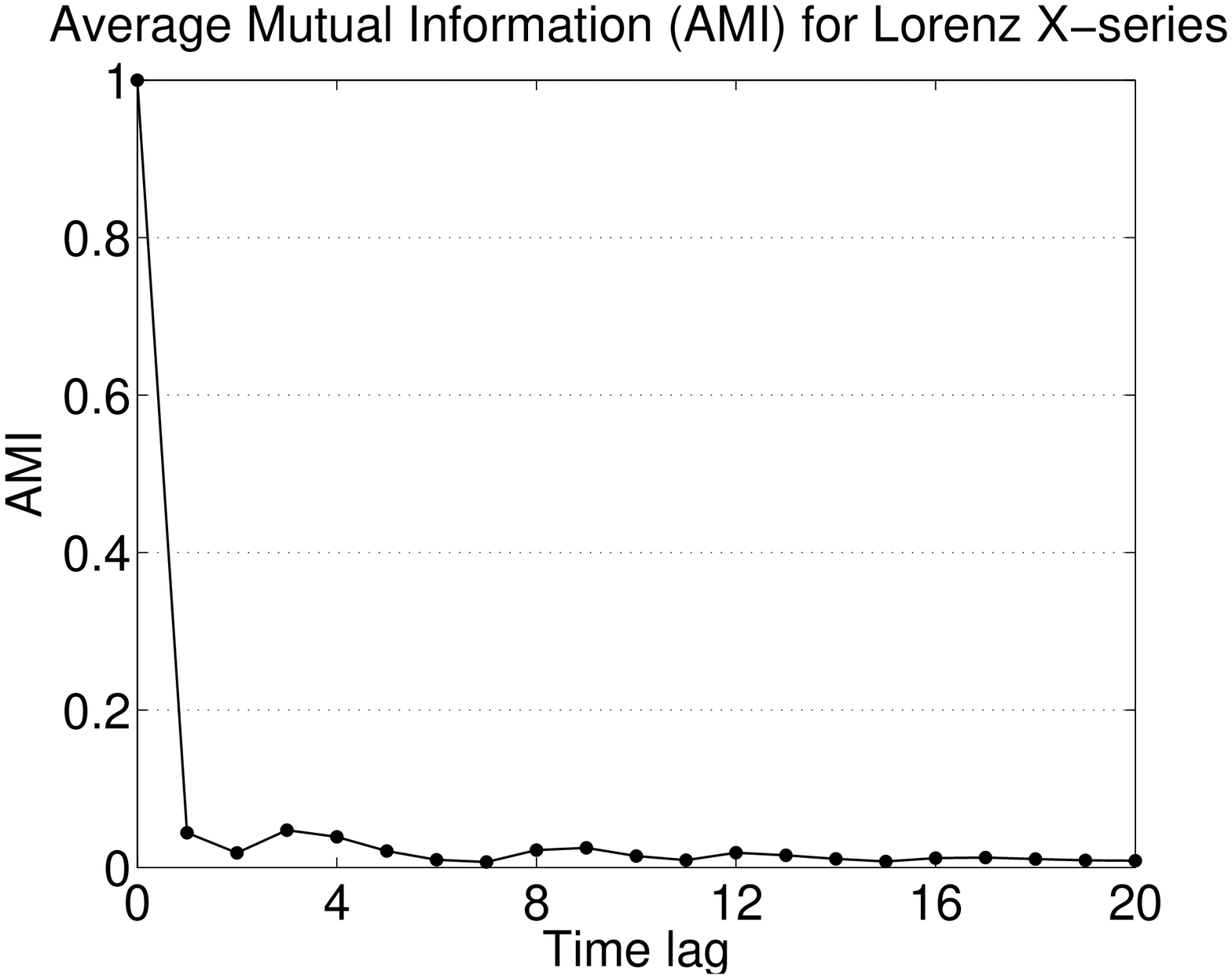,width=0.40\linewidth,clip=} & 
  \epsfig{file=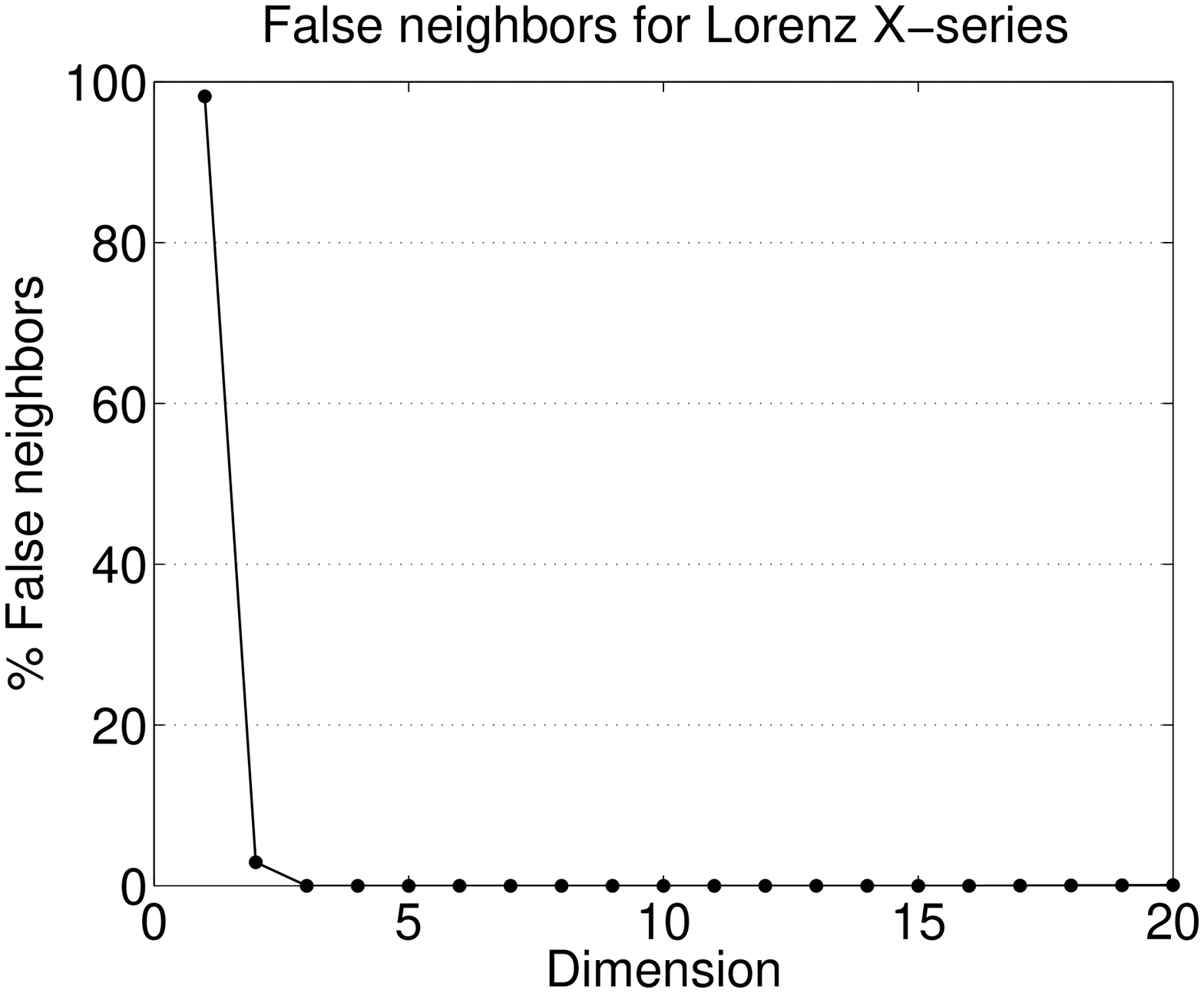,width=0.39\linewidth,clip=} \\
  \end{tabular}
  \vspace{-0.15in}
    \begin{verbatim}                   (a)                           (b)\end{verbatim}
\vspace{-0.1in}
\caption{(a) Average Mutual Information versus time lag plot used to fix the time lag $\tau$ for pure Lorenz X-series (b) \% of false neighborhood versus dimension plot for pure Lorenz X-series used to fix dimension d.}
\end{figure}

\begin{table}
\tbl{Number of false neighbors for pure Lorenz X-series with time lag $\tau = 1$.}
{\begin{tabular}{l|r|r|r|r|r} \hline
{\em Dimension} & 1 & 2 & 3 & 4 & 5 \\
\hline
{\em Number of false neighbors} & 8043 & 240 & 0 & 0 & 0 \\ \hline
\end{tabular} \label{ta:Lorenz_FN}}
\end{table}

Next we consider Lorenz X-series by adding Gaussian noise with noise levels 1\%, 5\% and 10\% separately. These X-series are smoothened with an Db-4 level 1 transformation. We then calculate average mutual information for various time lags and calculate \% of false neighbors for various dimensions. The time lags and dimensions obtained by above analysis are shown in Table 2.

\begin{table}
\tbl{Time lags and dimensions obtained for various Lorenz X-series having different noise levels.}
{\begin{tabular}{l|c|c|c|c|c} \hline
{\em Time series}	& {\em Time Lag}	& {\em Dimension}	&	& {\em Time Lag}	& {\em Dimension} \\
\hline
Smoothened Lorenz X(t) (no noise) &	1	& 3	& &	2	& 3 \\
Smoothened Lorenz X(t) with \ \ 1\% noise	& 1	& 4	& &	2	& 4 \\
Smoothened Lorenz X(t) with \ \ 5\% noise	& 1	& 4	& &	2	& 4 \\
Smoothened Lorenz X(t) with 10\% noise	& 1	& 5	& &	2	& 5 \\ \hline
\end{tabular} \label{ta:DimLag}}
\end{table}

\subsection{NASDAQ composite index time series}

Next we consider real financial NASDAQ composite index series. Fig. 5 shows the average mutual information for Db-4 levels 1 and 4 smoothened time series. As expected the correlations in level 1 series fall off slowly compared to the actual series. Also, for the level 4 series the approach to zero is even slower than level 1 series. We get $\tau = 6$ for level 1 time series and $\tau = 7$ for level 4 time series for the NASDAQ data.

\begin{figure}
\centering
  \begin{tabular}{c}
  \epsfig{file=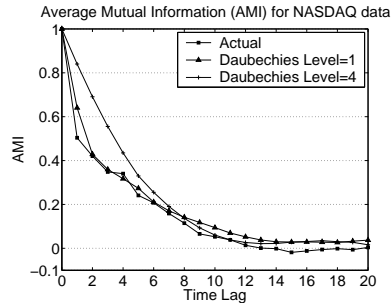,width=0.40\linewidth,clip=} \\
  \end{tabular}

\caption{Average Mutual Information versus time lag plot used to fix the time lag $\tau$ for Db-4 level 1 and 4 smooth NASDAQ time series.}
\end{figure}

We then determine the embedding dimension d using the number of false neighbor analysis. For Db-4 smoothened time series, the results for level 1 and level 4 series are shown in Table 3.

\begin{table}
\tbl{Time lags and dimensions for Db-4 transformed level 1 $(\tau=6)$ and level 4 $(\tau=7)$ smooth NASDAQ time series.}
{\begin{tabular}{l|c|c}                          \hline
 & {\em Time Lag $\tau$} & {\em Dimension d} \\  \hline
NASDAQ time series (Db-4 Level=1)	& 6	& 4    \\
NASDAQ time series (Db-4 Level=4)	& 7	& 6    \\  \hline
\end{tabular} \label{ta:lagNASDAQ}}
\end{table}

\section{Prediction using Lorenz Method}

We next consider a method due to Lorenz [10] for making predictions for a deterministic time series using time delay vector technique. We use the values of time delay $\tau$ and embedding dimension d obtained earlier. In order to make a prediction at $X_{i}$, the Lorenz method first constructs all available time delay vectors $y_{1}$, $y_{2}$, $y_{3}$ ... $y_{N}$ where $y_{N}$ is the last vector after which prediction is required. An embedding vector $y_{k}$ that matches the best with $y_{N}$ from all the previously known vectors is then searched for and on finding such a vector it is used as a predictor vector. For comparison of vectors, we choose a distance parameter called $\epsilon$ and see if vectors to be compared are closer than $\epsilon$. Initially we assign a small fraction of the size of the attractor as the starting value of $\epsilon$. If a vector is found within the limits of given $\epsilon$, that vector is used as a predictor. If more than one vectors are found, the average of the future values from each such vector is used as the estimated predicted value. However if no vector is found that is within the $\epsilon$ limit, the neighborhood size parameter $\epsilon$ is increased in an iterative manner till at least one such vector is found.

Very often the time series data of complex  systems is non-stationary and does not exhibit intermittency. In suh situations the Lorenz method of analogues is unlikely to provide a good prediction scheme. For this reason we have modified the Lorenz method in two ways as follows:

\begin{romanlist}[(ii)]
\item Instead of comparing a given vector to previous known vectors using absolute values of components, we propose to form {\em relative} vectors giving a better scope for matching the embedded vectors. While comparing a given embedded vector $y_{N}$ with a previously  known  embedded  vector  $y_{k}$,  the  components of $y_{k}$ are linearly shifted such that the first components of $y_{k}$ and $y_{N}$ match. The components of the two embedded vectors under consideration are therefore compared using their relative values rather than their absolute values. It ought to be clear that by doing this we are essentially comparing the pattern of variation in the components of the vectors. Hence, the scope of matching the vectors will be present in the entire earlier regime of the time series, unlike the situation where the absolute values of the components are matched. Consequently, we will have a much better possibility of capturing the dynamics and making predictions with the time series data. In view of this, we expect that the use of relative vectors will be very useful, especially for a non-stationary time series (e.g. NASDAQ series), having no intermittency.

\item Secondly we also require a minimum number of vectors \( (say \ 1, 3, 5) \) that need to be matched and make predictions using the average values. Out of different number of minimum vectors considered, we choose the one that gives the best prediction. 

\end{romanlist}

We first apply the above prediction schemes (using absolute and relative vectors) to Lorenz X-series without noise and with noise where the noise was  filtered using the Db-4 wavelet procedure described in Sec. 2. For both the prediction schemes, we increase the value of the size paramter $\epsilon$ till minimum of 1, 3 and 5 number of vectors are matched respectively. The best case prediction using absolute vectors and the best prediction using relative vectors are included. Results of the prediction are described next.

\subsection{Prediction for Lorenz X-series with and without noise}

In order to make prediction for Lorenz series, we choose 4 typical regions as shown in Fig. 1. Region 1 is a transition region in which the X value has a rapid fluctuation and then it tends to settle in a negative valued region. Region 2 is a transition region in which the X value has a rapid fluctuation and then it tends to settle in a positive valued region. Region 3 is a transition region in which the X value shifts from a positive valued region to that of a negative valued region. Region 4 is a region having rapidly varying X values. The 4 regions are chosen such that they have different characteristics to test the predictive capability of the prediction scheme.

Out of sample prediction for 50 points is made for above 4 regions for which the starting points are 360, 2950, 5135 and 7515. We calculate normalized mean square error (NMSE) defined in Eq. (4) as a measure of goodness of prediction:

\begin{equation}
NMSE = \frac{1}{N}  \frac{\sum^N_{i=1} [X_i^{calc}(m) - X_i^{given}(m)]^{2}  }
{variance \ of \ N \ data \ points}
\label{eqn:NMSE}
\end{equation}

where $X_{i}(m)$ are as per Eq. (1).

We find that NMSE as well as detailed predictions are in general better or comparable when the time lag  $\tau$ selected is 2 rather than 1. Table 4 shows NMSE values for the time series considered for the two time lags and different percentages of noise. As pointed out by Kulkarni et al [8], the value of $\tau$ appears to be crucial for capturing the dynamics on the attractor.

\begin{table}
\tbl{Out of sample prediction for 50 points starting at different regions for Lorenz X-series having different \% of noise level.}
{\begin{tabular}{ll|l|r|l|r|l|r|l|r} \hline
\multicolumn{2}{}{} & \multicolumn{4}{|c|}{Starting point 360} &
\multicolumn{4}{|c}{Starting point 2950} \\  \hline

 & & \multicolumn{2}{l|}{ABS vectors} & \multicolumn{2}{l|}{REL vectors} &
     \multicolumn{2}{l|}{ABS vectors} & \multicolumn{2}{l}{REL vectors}   \\ \hline

   Noise \% & $\tau$ & v & NMSE & v & NMSE & v & NMSE & v & NMSE \\  \hline \hline
0	 & 1	& 1	& 2.95 & 5 & 1.79	& 1	& 0.09 & 3 & 0.95 \\
   & 2	& 1	& 0.32 & 1 & 6.51 & 1	& 0.09 & 3 & 1.18	\\
1	 & 1	& 1	& 0.26 & 5 & 5.34	& 1	& 0.05 & 1 & 1.73	\\
	 & 2	& 1	& 0.25 & 3 & 0.73	& 1	& 0.05 & 3 & 0.45	\\
5	 & 1	& 1	& 2.60 & 5 & 4.50	& 1	& 2.12 & 1 & 0.04	\\
	 & 2	& 1	& 0.27 & 3 & 2.66	& 1	& 0.05 & 5 & 0.40	\\
10 & 1	& 1	& 1.64 & 5 & 2.98	& 5	& 2.05 & 1 & 0.26	\\
	 & 2	& 1	& 0.48 & 3 & 5.84	& 1	& 2.18 & 5 & 1.55	\\ \hline \hline

\multicolumn{2}{}{} & \multicolumn{4}{|c|}{Starting point 5135} &
\multicolumn{4}{|c}{Starting point 7515} \\  \hline



 & & \multicolumn{2}{l|}{ABS vectors} & \multicolumn{2}{l|}{REL vectors} &
     \multicolumn{2}{l|}{ABS vectors} & \multicolumn{2}{l}{REL vectors}   \\ \hline

0	 & 1	& 1	& 0.36	& 1	& 0.60	& 3	& 0.96 & 5 & 1.73 \\
   & 2	& 5	& 0.15	& 5	& 1.59	& 1	& 0.05 & 3 & 1.59 \\
1	 & 1	& 3	& 0.31	& 1	& 0.36	& 1	& 0.04 & 5 & 1.45 \\
	 & 2	& 1	& 0.36	& 1	& 0.36	& 1	& 0.04 & 3 & 0.64 \\
5	 & 1	& 1	& 0.36	& 3	& 1.65	& 5	& 0.23 & 3 & 1.95 \\
	 & 2	& 1	& 0.36	& 5	& 2.48	& 5	& 0.69 & 3 & 1.96 \\
10 & 1	& 1	& 0.36	& 1	& 1.12	& 5	& 0.40 & 3 & 1.32 \\
	 & 2	& 3	& 0.33	& 5	& 0.89	& 1	& 0.04 & 1 & 1.18 \\
\hline
\end{tabular} \label{ta:LorenzPred}}
\end{table}

As illustration we give detailed results only for Lorenz X-series having 5\% Gaussian noise where the noise was filtered using Db-4 wavelet procedure. Fig. 6 shows out of sample prediction for 50 points beginning at point numbers 5135 and 7515 which correpond to the best and the worst cases resepctively out of the 4 cases considered.

\begin{figure}
\centering

  \begin{tabular}{cc}
  \epsfig{file=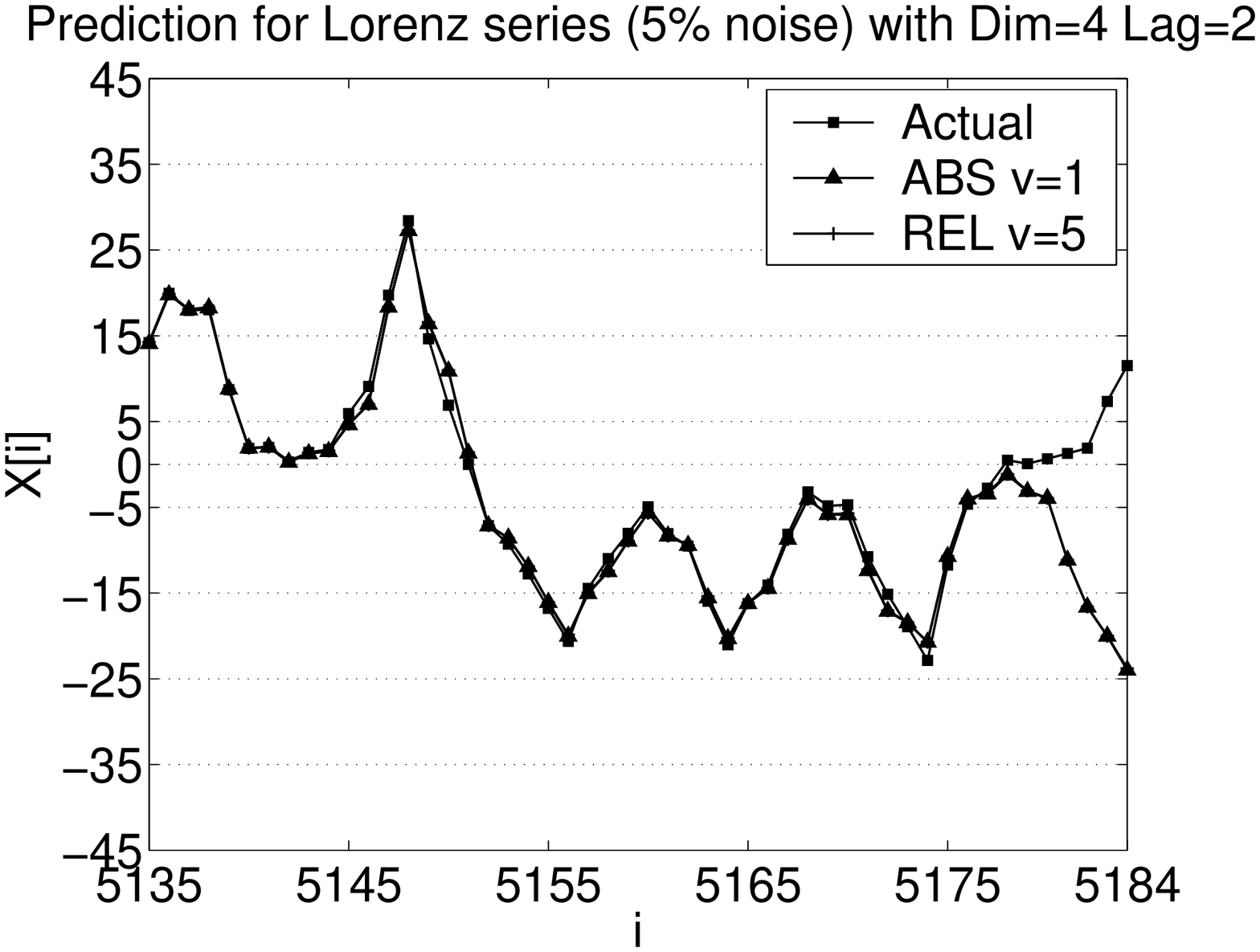,width=0.40\linewidth,clip=} & 
  \epsfig{file=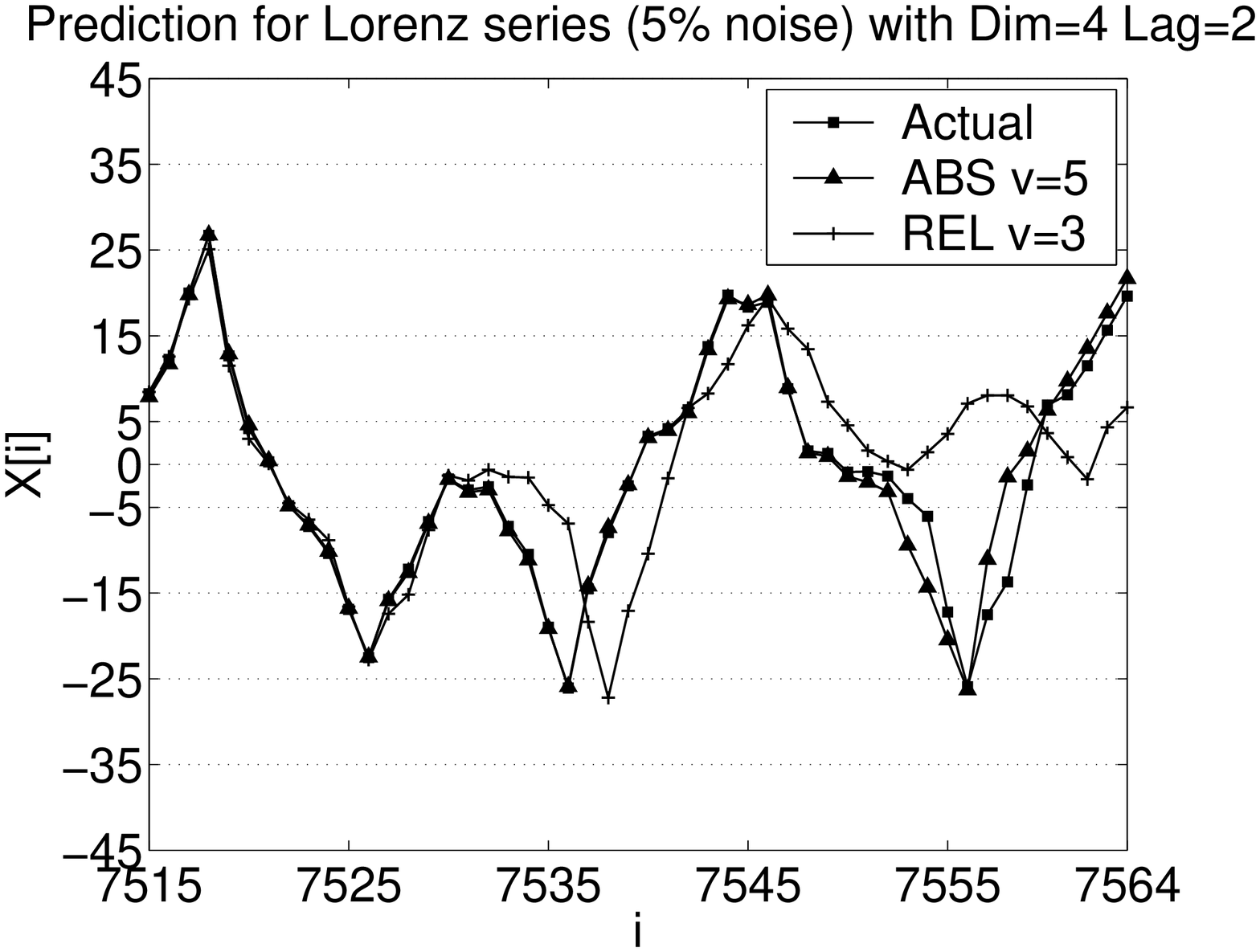,width=0.40\linewidth,clip=} \\
  \end{tabular}
  \vspace{-0.15in}
    \begin{verbatim}                   (a)                          (b)\end{verbatim}

\vspace{-0.15in}
\caption{Out of sample prediction of time series for 50 points using smoothened $(\tau=2, dim=4)$ Lorenz X-series (with 5\% noise). A dynamic prediction that uses successively predicted vectors is made. The best of the prediction (out of vectors=1, 3 and 5) corresponding to absolute (ABS) and relative (REL) methods have been used. The legend also indicates the number of minimum vectors (labelled by v) that gives the best prediction. The prediction is made for points starting at (a) 5135 and (b) 7515.}
\end{figure}

We observe that the quality of predictions vary for the four different starting points considered in the series. However, in most cases, predictions made with absolute values of the delay vector components are superior to the ones made with relative values of the components of the delay vectors.

\subsection{Prediction for NASDAQ composite index Db-4 level=1 smooth time series}

For NASDAQ series, we again choose 4 typical regions (as shown in Fig. 2 (a)) as follows. Region 1 is a transition region in which the X value has a rapid variation and is in the first half of the series. Region 2 corresponds to the overall peak of the entire data set. Region 3 has a local minimum of the X values. Region 4 has fluctuations of the X values towards the end of the data set. The 4 regions are chosen such that they have different characteristics to test the predictive capability of the scheme.

Out of sample prediction for 20 points are made for above 4 regions for which the starting points are 790, 1016, 1288 and 1825. Normalized mean square error (NMSE) is calculated to infer the goodness of prediction schemes.

The NMSE errors and the minimum and maximum errors for the predictions are shown in Table 5. Note that for this time series the values of NMSE are lower (or comparable) when the predictions are made using relative vectors.

\begin{table}
\tbl{NMSE and minimum, maximum deviations between calculated and actual data values for predictions (for NASDAQ Db-4 level 1 time series) starting at points (a) 790, (b) 1016, (c) 1288 and (d) 1825 respectively using absolute (ABS) and relative (REL) methods.}
{\begin{tabular}{l|l|r|r|r} \hline


 Starting point	  & No. of vectors	& NMSE	& Min. Diff.	& Max. Diff.  \\ \hline

790	  & ABS v=5	& 10.71	& -55.41	& -308.13 \\
	    & REL v=3	& 1.68	& -3.11	  & 216.03  \\
1016	& ABS v=3	& 28.80	& -395.91	& -873.21 \\
	    & REL v=3	& 1.48	& 0.01	  & 6.60    \\
1288	& ABS v=5	& 1.18	& -1.97	  & -375.63 \\
	    & REL v=1	& 1.21	& -6.00	  & -365.62 \\
1825	& ABS v=5	& 0.79	& -2.61	  & -92.58  \\
	    & REL v=1	& 0.63	& 5.73	  & -90.66  \\

\hline
\end{tabular} \label{ta:NMSEvect1}}
\end{table}

In Fig. 7 the predicted time series is compared starting at different locations in the time series with the actual time series starting at point numbers 1016 and 1825 which correspond to the best and the worst cases respectively out of the 4 cases considered. 

\begin{figure}
\centering

  \begin{tabular}{cc}
  \epsfig{file=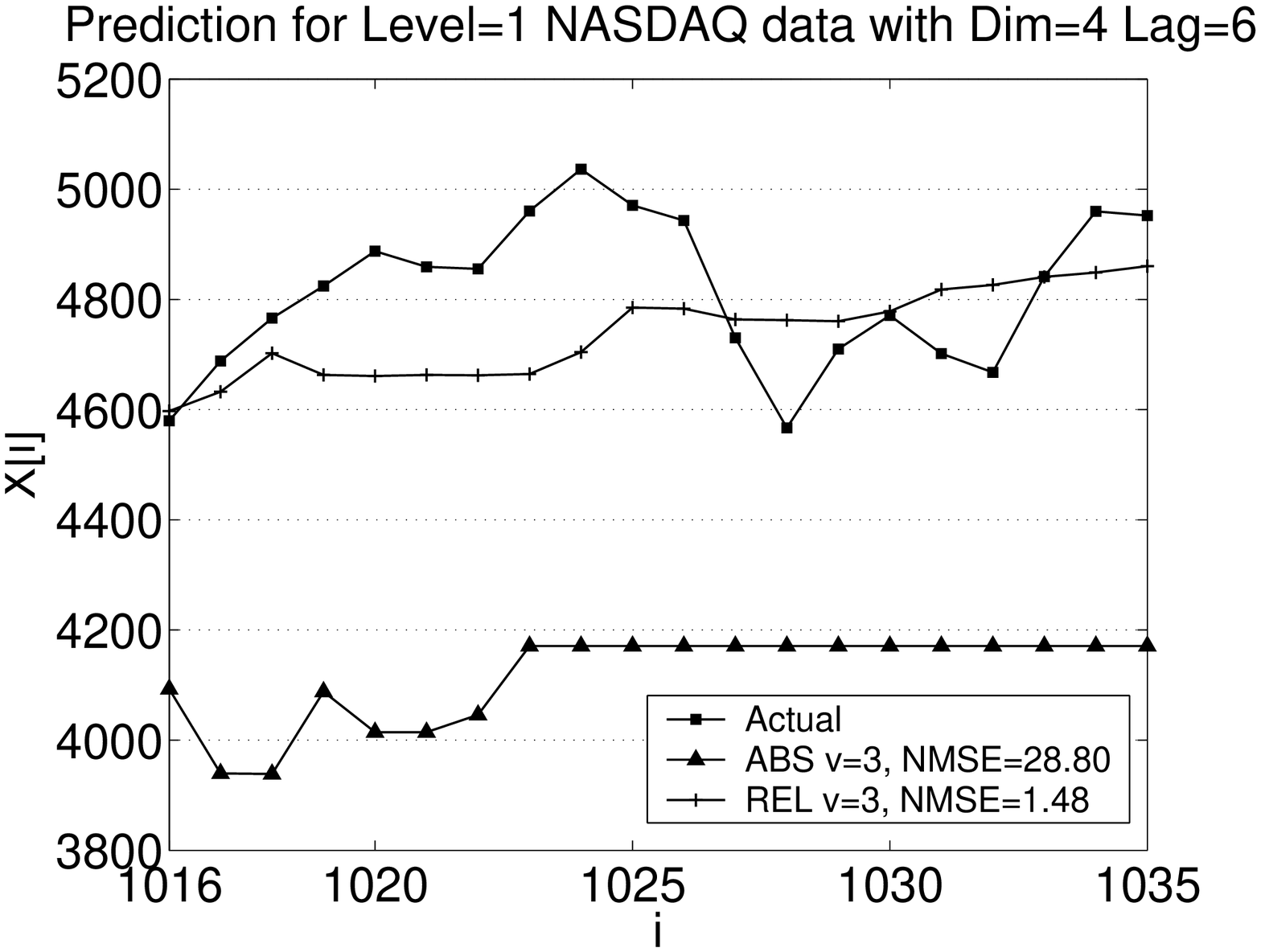,width=0.40\linewidth,clip=} & 
  \epsfig{file=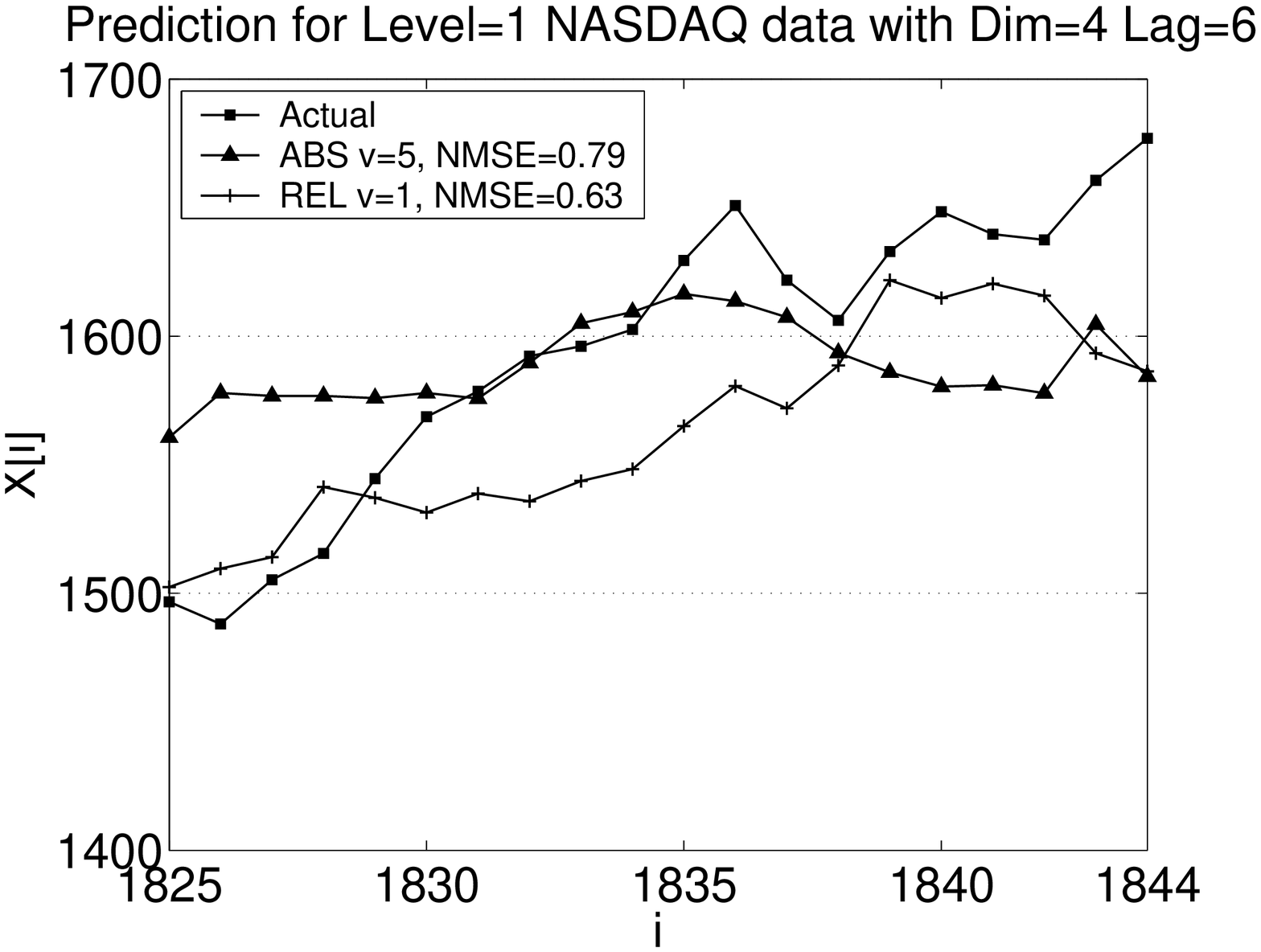,width=0.40\linewidth,clip=} \\
  \end{tabular}
  \vspace{-0.15in}
    \begin{verbatim}                   (a)                          (b)\end{verbatim}

\vspace{-0.15in}
\caption{Out of sample prediction of time series for 20 points using NASDAQ level 1 smoothened time series. A dynamic prediction that uses successively predicted vectors is made. The best of the prediction (out of vectors=1, 3 and 5) corresponding to absolute (ABS) and relative (REL) methods have been used. The legend also indicates the number of minimum vectors (labelled by v) that gives the best prediction. The prediction is made for points starting at points (a) 1016 and (b) 1825.}
\end{figure}

\subsection{Prediction for NASDAQ composite index Db-4 level=4 smooth time series}

We next consider NASDAQ composite index series smoothened using Db-4 level=4 transformation.

Out of sample predication for 20 points is made for above 4 regions for which the starting points are 790, 1016, 1288 and 1825. Normalized mean square error (NMSE) is calculated for all predictions made to infer about the goodness of prediction between the actual and calculated values.

The NMSE errors and the minimum and maximum errors for the predictions are shown in Table 6. We find that NMSE as well as detailed predictions are in general better when the relative vectors are used as compared to the absolute vectors.

\begin{table}
\tbl{NMSE and minimum, maximum deviations between calculated and actual data values for predictions (for NASDAQ Db-4 level 4 smooth time series) starting at points (a) 790, (b) 1016, (c) 1288 and (d) 1825 respectively using absolute (ABS) and relative (REL) methods.}
{\begin{tabular}{l|l|r|r|r} \hline


 Starting point	  & No. of vectors	& NMSE	& Min. Diff.	& Max. Diff.  \\ \hline

790	   & ABS v=5	& 85.64	& -127.39	& -191.36 \\
	     & REL v=3	& 0.50	& 0.73	  & 31.01   \\
1016	 & ABS v=1	& 58.62	& -707.11	& 1107.91 \\
	     & REL v=5	& 3.70	&  25.42	& 461.97  \\
1288	 & ABS v=5	& 22.15	& 296.90	& 699.20  \\
	     & REL v=1	& 2.58	& -5.09	  & 350.82  \\
1825	 & ABS v=3	& 0.61	& 1.76	  & -62.02  \\
	     & REL v=3	& 1.50	& -2.31	  & -74.02  \\

\hline
\end{tabular} \label{ta:NMSEvect2}}
\end{table}

In Fig. 8 the predicted time series is compared starting at different locations in the time series with the actual time series starting at point numbers 790 and 1825 which correspond to the best and the worst cases respectively out of the 4 cases considered. 

\begin{figure}
\centering

  \begin{tabular}{cc}
  \epsfig{file=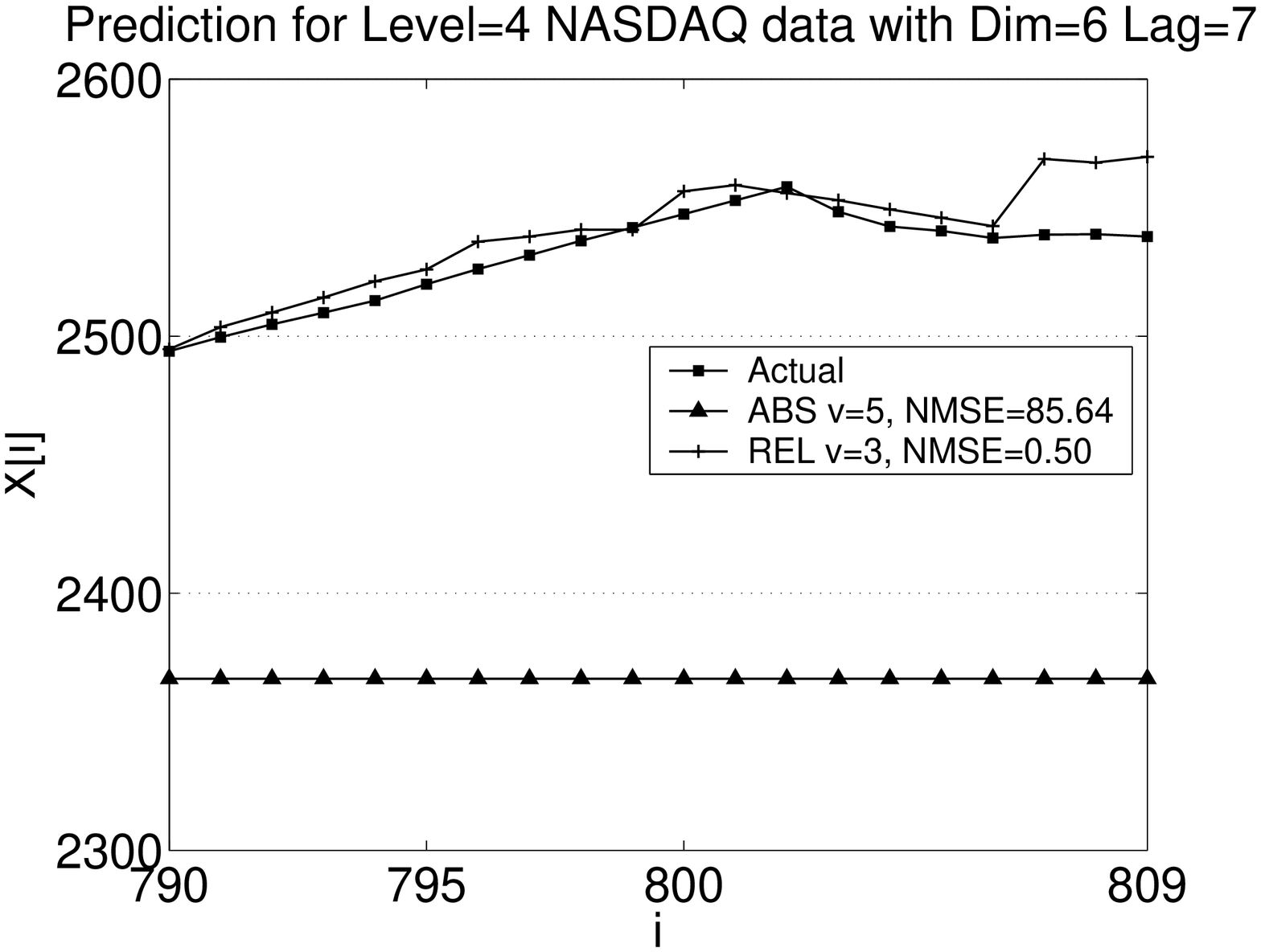,width=0.40\linewidth,clip=} & 
  \epsfig{file=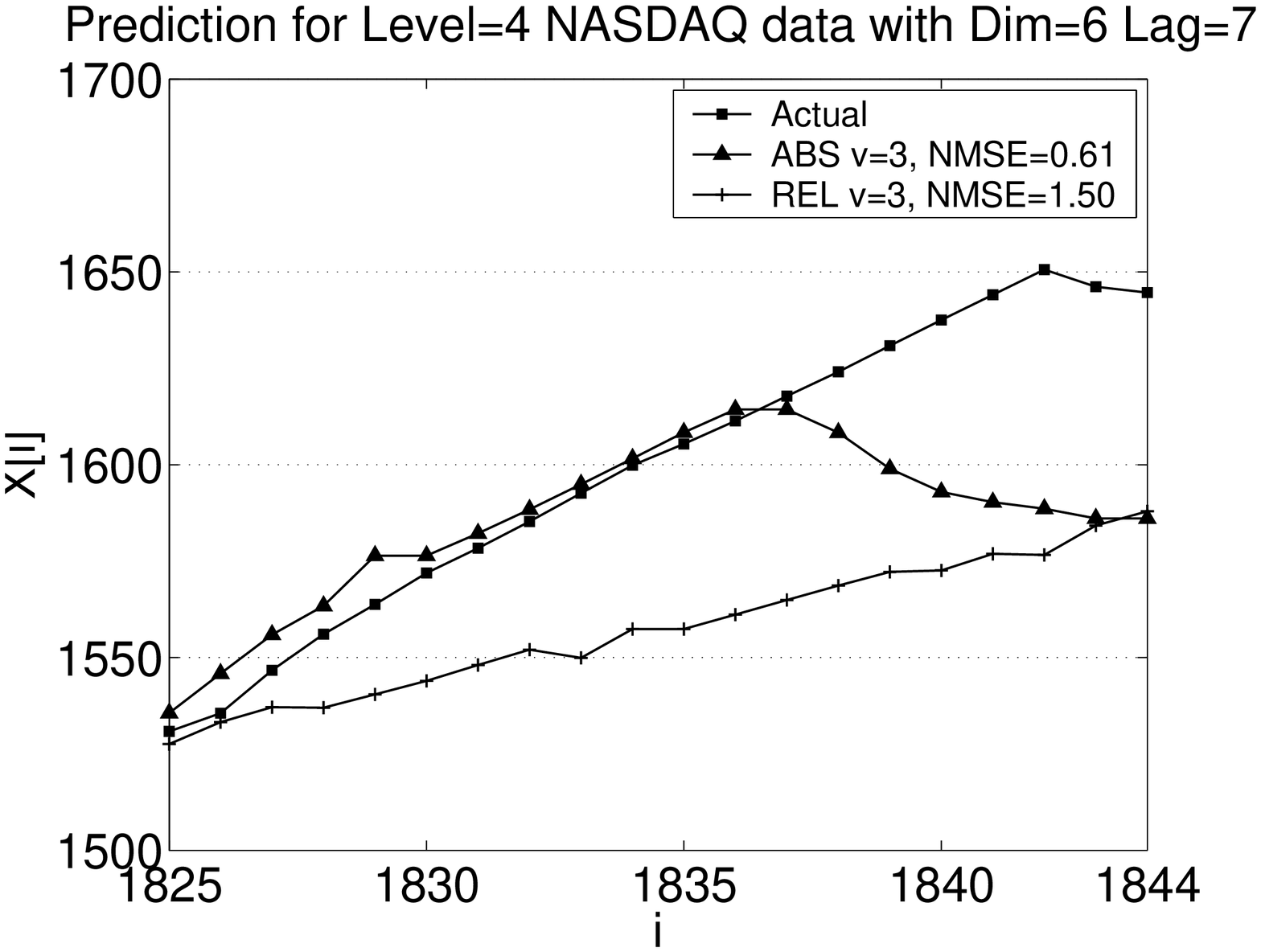,width=0.40\linewidth,clip=} \\
  \end{tabular}
  \vspace{-0.15in}
    \begin{verbatim}                  (a)                           (b)\end{verbatim}

\vspace{-0.15in}
\caption{Out of sample prediction of time series for 20 points using NASDAQ level 4 time series. A dynamic prediction that uses successively predicted vectors is made. The best of the prediction (out of vectors=1, 3 and 5) corresponding to absolute (ABS) and relative (REL) methods have been used. The legend also indicates the number of minimum vectors (labelled by v) that gives the best prediction. The prediction is made for points starting at points (a) 790 and (d) 1825.}
\end{figure}

It is evident that the prediction of time series using modified Lorenz method gives better results as compared to the Lorenz method.

From Tables 5 and 6 it is interesting to note that for the NASDAQ time series, the Lorenz method does require the use of more than one vectors to achieve better predictions.

We note here an interesting connection between the prediction quality and the size parameter $\epsilon$. As discussed earlier, in the Lorenz prediction method, the value of the neighborhood size parameter $\epsilon$ is increased till a given number of matching embedded vectors (say 1, 3, 5 ...) are found. It is therefore interesting to see the role of the size parameter in the quality of prediction. The variation of size parameter $\epsilon$ is shown in Fig. 9 (a) for level=1 NASDAQ series (having attractor size 837.57) for prediction of 20 points starting at point 1016. Fig. 9 (b) shows variation of size parameter $\epsilon$ for level=4 NASDAQ series (having attractor size 835.14) for prediction of 20 points starting at point 790.

\begin{figure}
\centering

  \begin{tabular}{cc}
  \epsfig{file=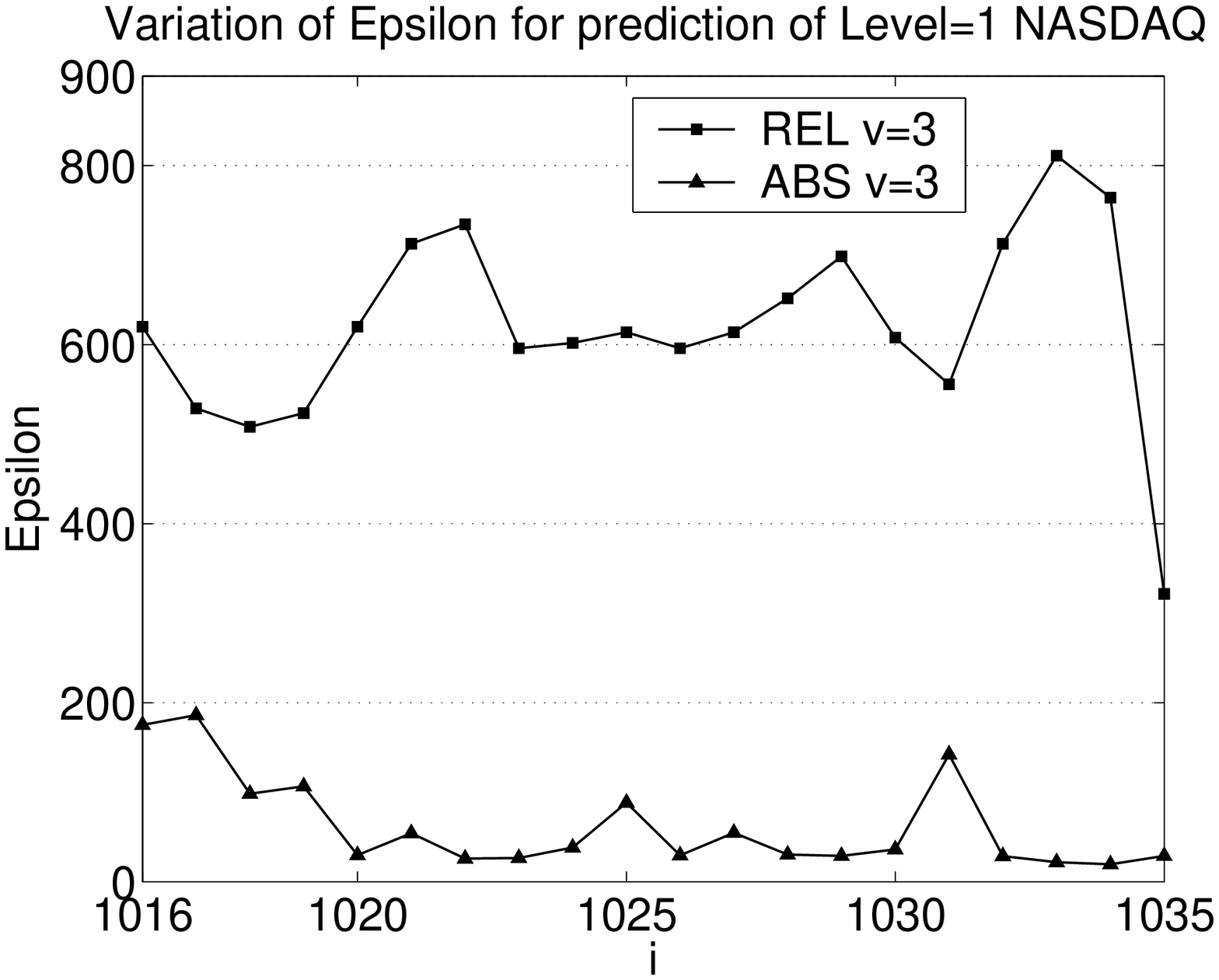,width=0.40\linewidth,clip=} & 
  \epsfig{file=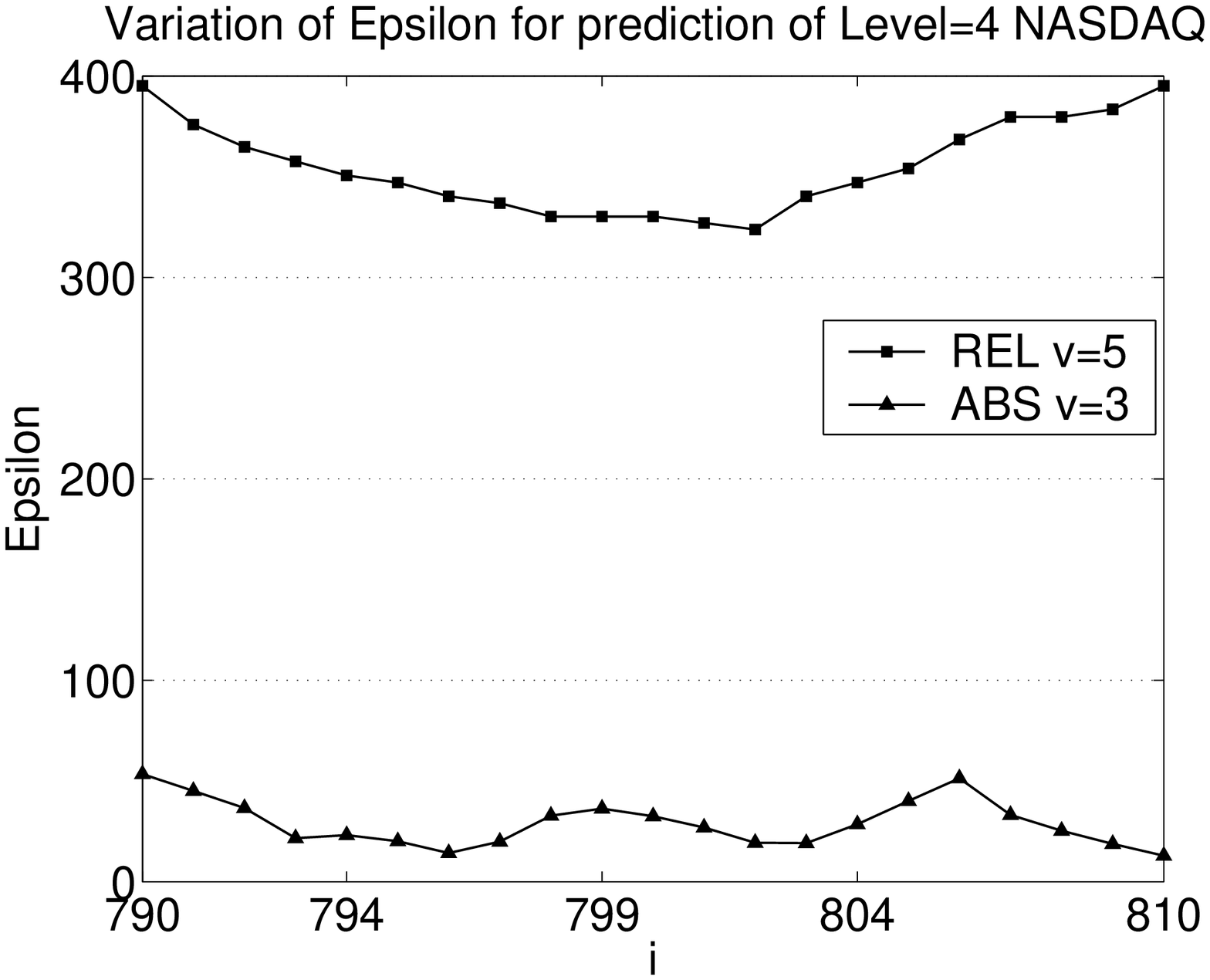,width=0.40\linewidth,clip=} \\
  \end{tabular}
  \vspace{-0.15in}
    \begin{verbatim}                   (a)                           (b)\end{verbatim}

\vspace{-0.15in}
\caption{Variation of size parameter $\epsilon$ for (a) Db-4 level=1 smooth time series for prediction of 20 points starting at point 1016 and (b) Db-4 level=4 smooth time series for prediction of 20 points starting at point 790. The detailed predictions are shown earlier in Fig. 7 (a) and 8 (a) respectively.}
\end{figure}

It is evident from Fig. 9 that the prediction of time series is in general good whenever vectors from the previously known data are found corresponding to a reasonably small value (compared to the size of the attractor) of $\epsilon$ parameter. If the $\epsilon$ parameter has to be increased substantially, the quality of prediction is in general adversely affected. It is therefore interesting to note that monitoring of $\epsilon$ value gives us an indirect clue regarding the goodness of prediction.

\section{Analysis of Fluctuations}

As is evident from Figs. 1 and 2 in Sec. 2, fluctuations in the time series data can be obtained by subtracting the trend values ($X_{i}(m)$ in Eq. 1) from the actual data. In this section we analyze the fluctuation properties of NASDAQ level 1 and level 4 series. More precisely, for these fluctuations we obtain

\begin{enumerate}
\item Probability distribution and its cumulants
\item Auto-correlation function (ACF)
\end{enumerate}

\subsection{Probability Distribution Function (PDF)}

We show in Fig. 10 the PDF of standardized variables

\begin{equation}
S_i = \frac{[X_i(f) - \overline{X}(f)]} {\sigma_f}
\label{eqn:StandardVar}
\end{equation}

for the two data sets (where $\sigma_f$ is the standard deviation for the fluctuations X(f) series).

\begin{figure}
\centering

  \begin{tabular}{cc}
  \epsfig{file=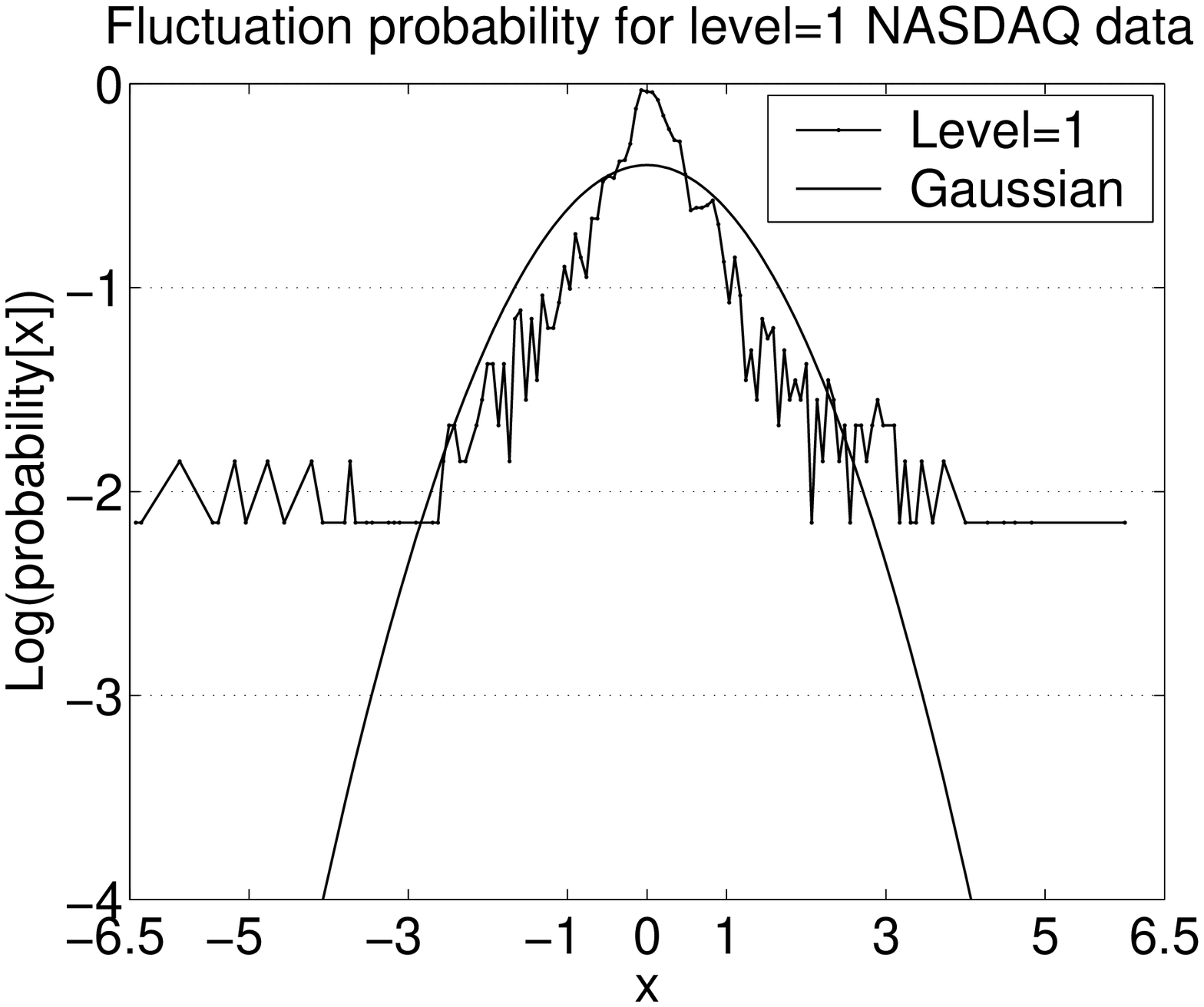,width=0.40\linewidth,clip=} & 
  \epsfig{file=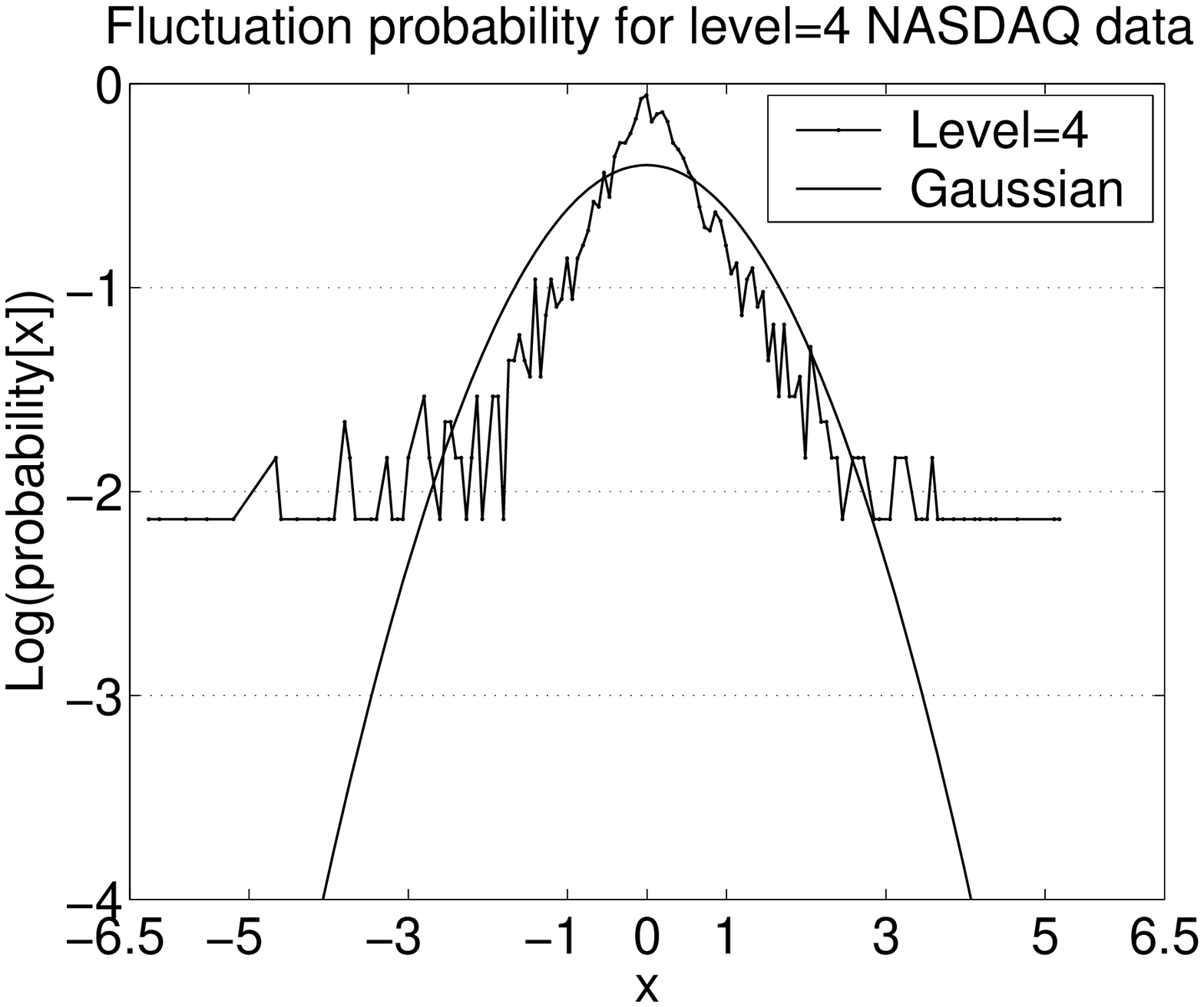,width=0.40\linewidth,clip=} \\
  \end{tabular}
  \vspace{-0.15in}
    \begin{verbatim}                   (a)                           (b)\end{verbatim}

\vspace{-0.15in}
\caption{Log-linear plots for the fluctuation probability distributions for (a) Db-4 level=1 NASDAQ data and (b) Db-4 level=4 NASDAQ data.}
\end{figure}

Note that log of the probabilities are plotted so that the tails of the distributions are easily distinguished from those of a Gaussian distribution. The PDF of NASDAQ fluctuations show significant departures from the Gaussian form. They exhibit fatter tails and are peaked more in the center relative to the Gaussian PDF. We also give in Table 7 values of cumulants, namely skewness $(\gamma_1)$ and kurtosis $(\gamma_2)$ for the two fluctuation data sets.

\begin{table}
\tbl{Skewness($\gamma_1$) and Kurtosis ($\gamma_2$) for the NASDAQ level=1 fluctuation series and NASDAQ level=4 fluctuation series.}
{\begin{tabular}{l|r|r} \hline

                            & Skewness $\gamma_1$	& Kurtosis $\gamma_2$ \\ \hline

\hspace{0.2in} NASDAQ level=1 fluctuations &	-0.409	& 9.345 \\
\hspace{0.2in} NASDAQ level=4 fluctuations &	-0.597	& 9.104 \\   \hline
\end{tabular} \label{ta:GammaKurtosis}}
\end{table}

It may be recalled that for Gaussian random noise, $\gamma_1$=0 and $\gamma_2$=0. Non-negligible negative values of $\gamma_1$ that we find imply that the probability distributions are asymmetric with greater weight for values less than the mean. Large positive kurtosis in probability distributions is an indication that the peak is higher and the tails are fatter than those for a Gaussian distribution. These features are clearly seen in Fig. 11.

\subsection{Auto-Correlation Function (ACF)}

We also evaluate ACF for fluctuations and absolute value of fluctuations. These are shown in Fig. 11 (for fluctuations) and Fig. 12 (for absolute values of fluctuations).

\begin{figure}
\centering
  \begin{tabular}{cc}
  \epsfig{file=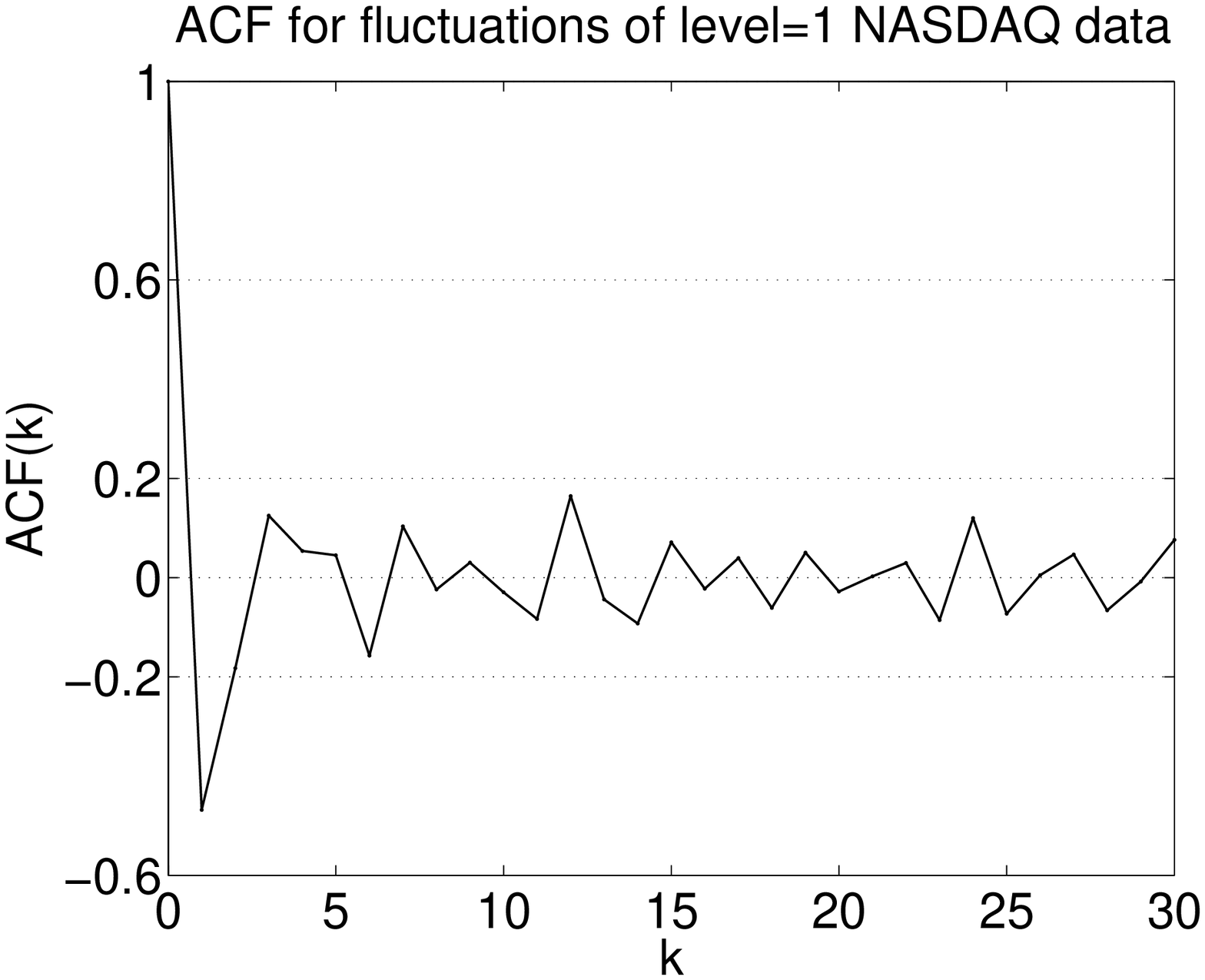,width=0.40\linewidth,clip=} & 
  \epsfig{file=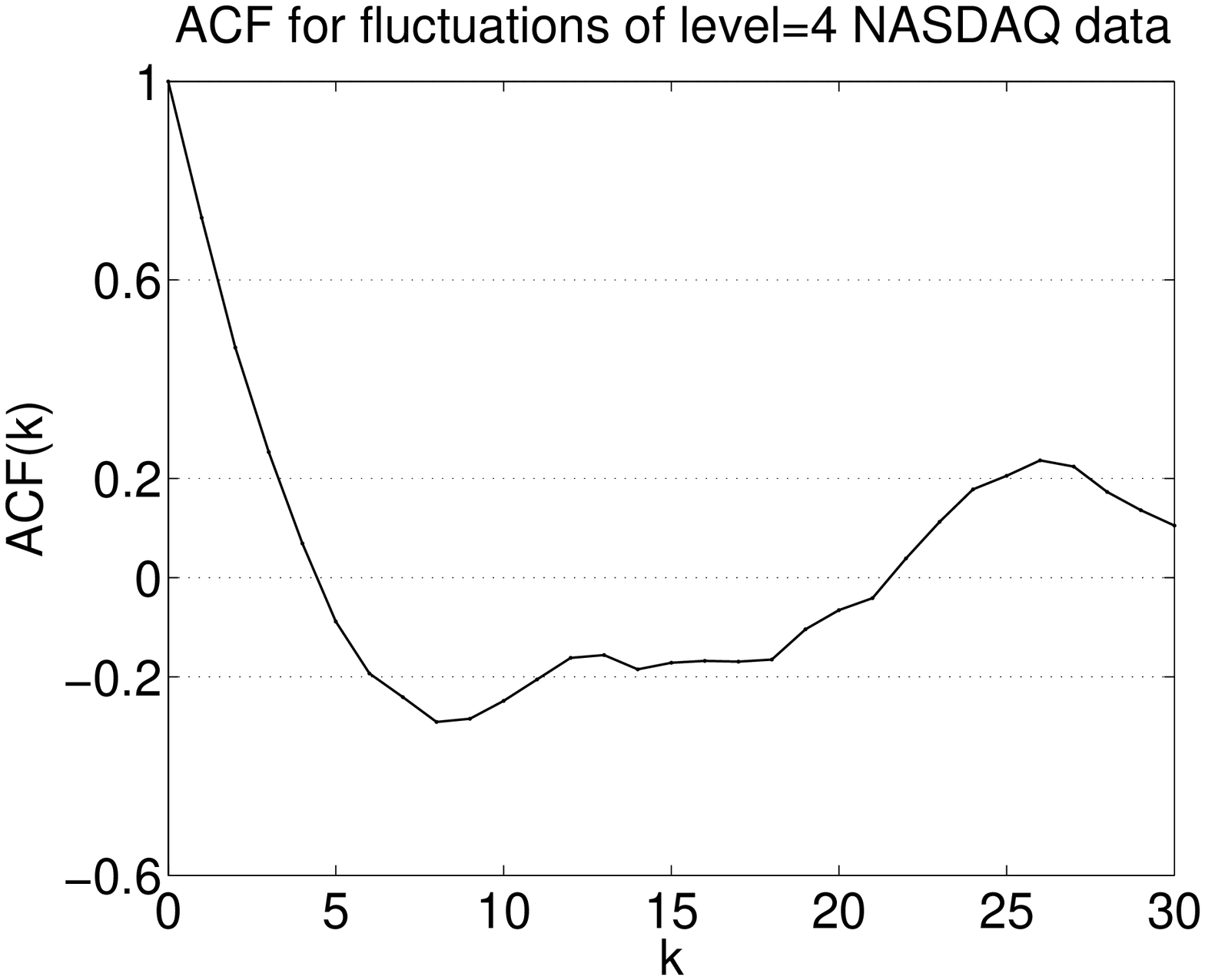,width=0.40\linewidth,clip=} \\
  \end{tabular}
  \vspace{-0.15in}
    \begin{verbatim}                   (a)                           (b)\end{verbatim}

\vspace{-0.15in}
\caption{Auto-Correlation Functions of fluctuations for (a) Db-4 level=1 NASDAQ data and (b) Db-4 level=4 NASDAQ data.}
\end{figure}

\begin{figure}
\centering
  \begin{tabular}{cc}
  \epsfig{file=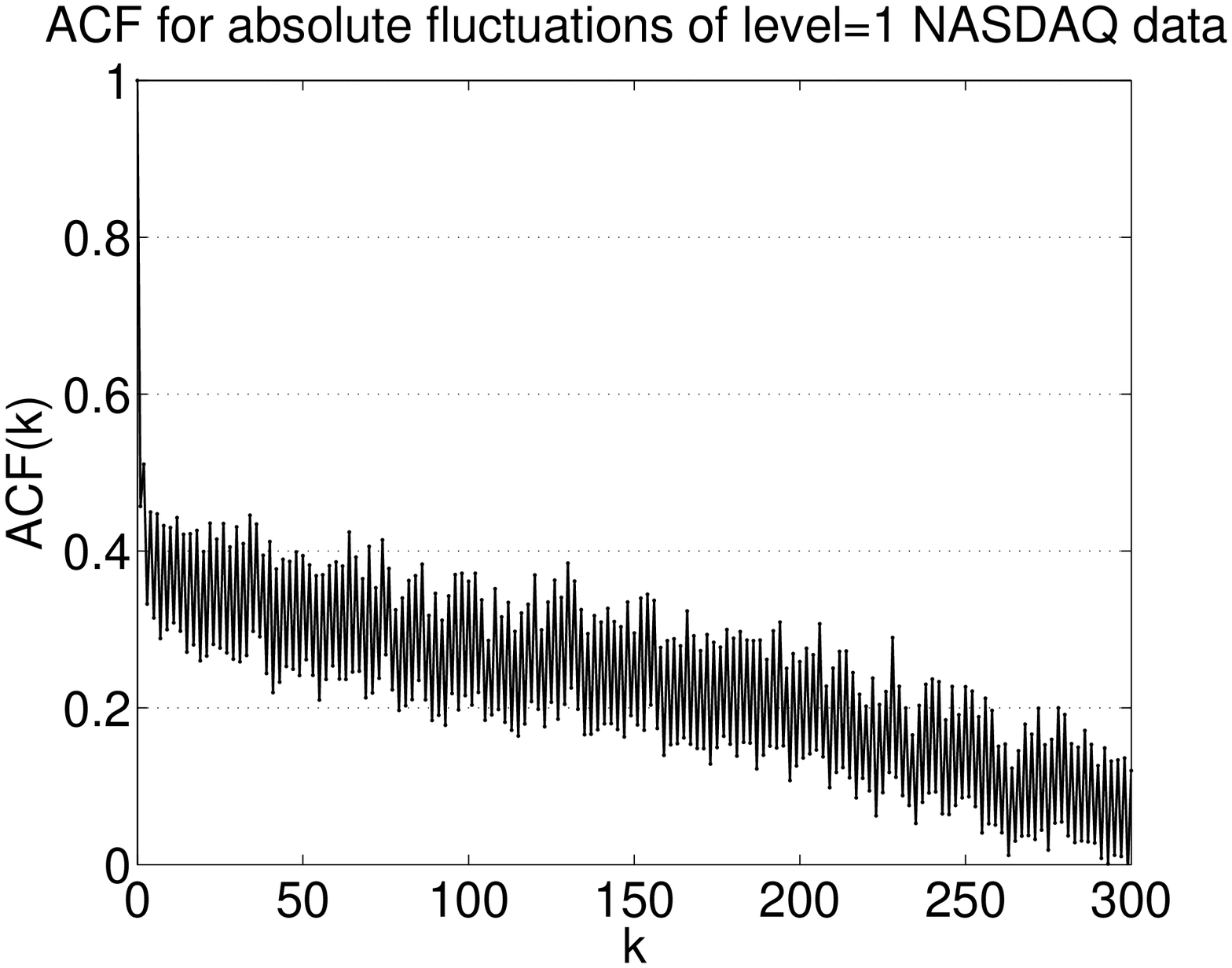,width=0.40\linewidth,clip=} & 
  \epsfig{file=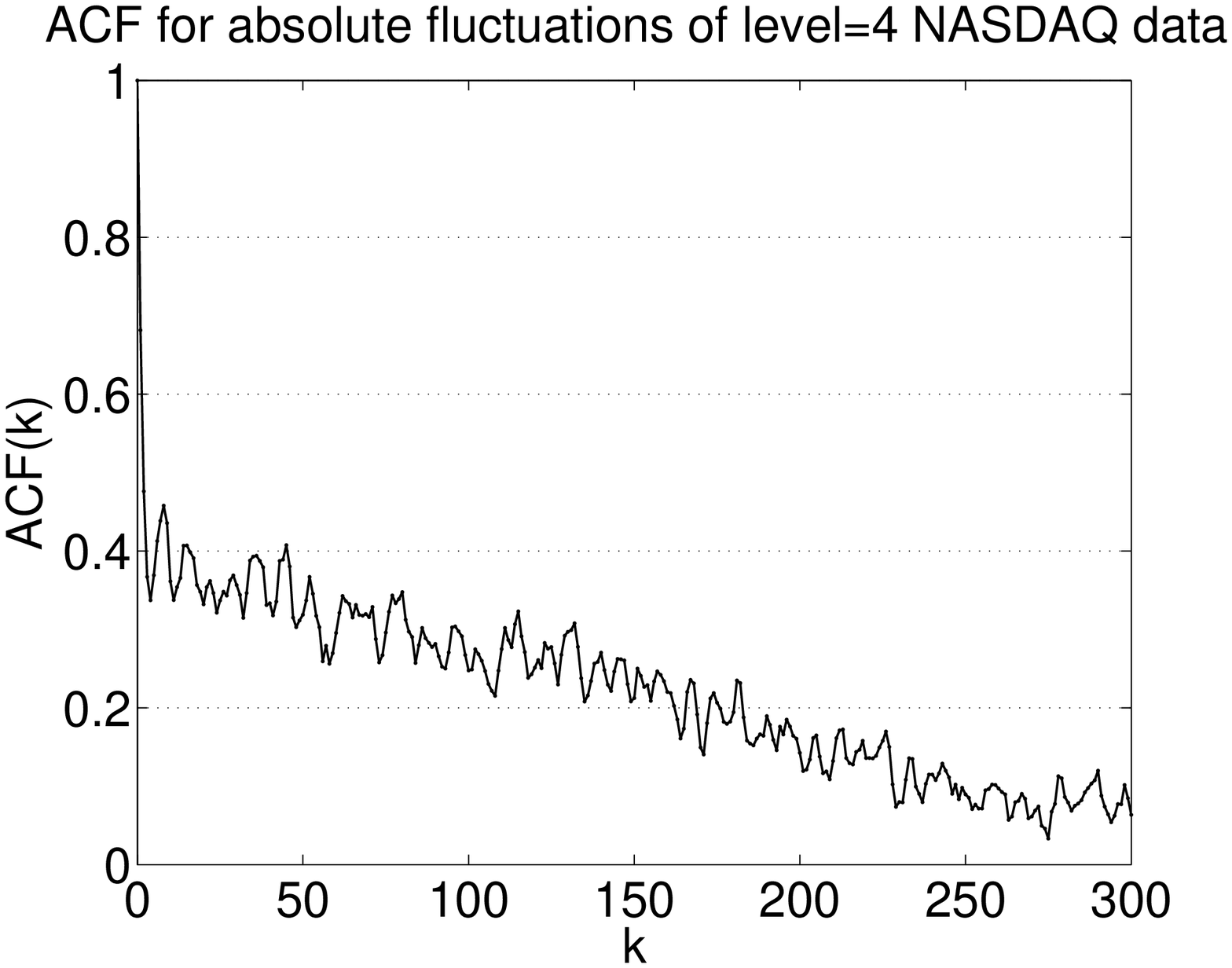,width=0.40\linewidth,clip=} \\
  \end{tabular}
  \vspace{-0.15in}
    \begin{verbatim}                   (a)                           (b)\end{verbatim}

\vspace{-0.15in}
\caption{Auto-Correlation Functions of absolute values of fluctuations for (a) Db-4 level=1 NASDAQ data and (b) Db-4 level=4 NASDAQ data.}
\end{figure}

We see that the ACF for the fluctuations drop to small values as a function of time lag. We further find that the fluctuations are not correlated beyond a few days. Note that by contrast absolute values of fluctuations remain correlated for very long time. This means that although the fluctuations do not have significant (binary) correlations, they are not independent.

\section{Summary}
Considering Lorenz X-series having 8192 data points (with and without Gaussian noise) and NASDAQ composite index time series having 2048 data points, we have smoothened the time series (at level 1 for Lorenz series and levels 1 and 4 for NASDAQ series) using a wavelet (Db-4) transformation. The low-pass wavelet coefficients provide a representation (in transformed space) for the deterministic trend of the time series.

We have then reconstructed the state space for these smoothened time series and obtained time delay vectors using the appropriate time lags $\tau$ and dimension d. The trend series can also be modelled in the Artificial Neural Network (ANN) framework. We have not done so. We use a method proposed by Lorenz [10] for prediction purposes. We also propose a modification in the method by using {\em relative} vectors for making predictions. We find that the Lorenz method with absolute vectors is good for predictions of Lorenz X-series (having intermittency and stationary nature of time series). However, for the NASDAQ financial time series (having a non-stationary nature of the time series) relative vectors give better prediction. We also use various number of vectors to be matched (1, 3 and 5) and choose the one that gives the best result. Our observation that lower values for the size parameter $\epsilon$ correspond to better predictions can serve as a handle in monitoring the quality of predictions.

Analysis of fluctuation properties shows that for the NASDAQ data the probability distribution is non-Gaussian. They are asymmetric and have fat tails. The fluctuations are weakly correlated in time but are not independent.

In conclusion we would like to state we have modeled the trend in the time series through low pass wavelet coefficients shown in Figs. 3 (a) and 3 (b) and the fluctuations in the time series through the statistical properties described in Sec. 5. Thus we have a comprehensive model for the time series.

\end{document}